\documentstyle[11pt,axodraw, ijmpal-1]{article}
\begin{document}
\newcounter{myfn}[page]
\renewcommand{\thefootnote}{\fnsymbol{footnote}}
\newcommand{\myfootnote}[1]{\setcounter{footnote}{\value{footnote}}%
\footnote{#1}\stepcounter{footnote}}
\renewcommand{\theequation}{\thesection.\arabic{equation}}
\newcounter{saveeqn}
\newcommand{\add}{\addtocounter{equation}{1}}
\newcommand{\alpheqn}{\setcounter{saveeqn}{\value{equation}}%
\setcounter{equation}{0}%
\renewcommand{\theequation}{\mbox{\thesection.\arabic{saveeqn}{\alph{equation}}}}}
\newcommand{\reseteqn}{\setcounter{equation}{\value{saveeqn}}%
\renewcommand{\theequation}{\thesection.\arabic{equation}}}
\newenvironment{nedalph}{\add\alpheqn\begin{eqnarray}}{\end{eqnarray}\reseteqn}
\newsavebox{\PSLASH}
\sbox{\PSLASH}{$p$\hspace{-1.8mm}/}
\newcommand{\PS}{\usebox{\PSLASH}}
\newsavebox{\ASLASH}
\sbox{\ASLASH}{$A$\hspace{-1.8mm}/}
\newcommand{\AS}{\usebox{\ASLASH}}
\newsavebox{\KSLASH}
\sbox{\KSLASH}{$k$\hspace{-1.8mm}/}
\newcommand{\KS}{\usebox{\KSLASH}}
\newsavebox{\LSLASH}
\sbox{\LSLASH}{$\ell$\hspace{-1.8mm}/}
\newcommand{\LS}{\usebox{\LSLASH}}
\newsavebox{\SSLASH}
\sbox{\SSLASH}{$s$\hspace{-1.8mm}/}
\newcommand{\SS}{\usebox{\SSLASH}}
\newsavebox{\DSLASH}
\sbox{\DSLASH}{$D$\hspace{-2.4mm}/}
\newcommand{\DS}{\usebox{\DSLASH}}
\setcounter{footnote}{0}
\renewcommand{\thefootnote}{\alph{footnote}}
\thispagestyle{empty}
\begin{flushright}
IPM/P-2000/009
\par
hep-th/0002143
\end{flushright}
\vspace{0.5cm}
\begin{center}
{\Large\bf{Axial Anomaly in Noncommutative QED on $R^{4}$}}\\
\vspace{1cm}
{\bf Farhad Ardalan}\footnote{\normalsize{Electronic address: ardalan@theory.ipm.ac.ir}}
\hspace{0.2cm}
and \hspace{0.2cm}{\bf N\'eda Sadooghi}\footnote{\normalsize{Electronic address: sadooghi@theory.ipm.ac.ir}}  \\
\vspace{0.5cm}
{\sl Institute for Studies in Theoretical Physics and Mathematics (IPM)}\\
{\sl{School of Physics, P.O. Box 19395-5531, Tehran-Iran}}\\
and\\
{\sl Department of Physics, Sharif University of Technology}\\
{\sl P.O. Box 11365-9161, Tehran-Iran}
\end{center}
\vspace{1cm}
\begin{center}
{\bf {Abstract}}
\end{center}
\begin{quote}
The axial anomaly of the noncommutative U(1) gauge theory is calculated by a number of methods and compared with the commutative one. It is found to be given by the corresponding Chern class. 
\end{quote}
\par\noindent
\vskip4cm

\newpage
\setcounter{page}{1}
\section{Introduction}\noindent
Quantum field theory on noncommutative spaces\cite{connes,witten} has been studied intensively in the last few months:  In Ref.\cite{minwalla} certain properties of the perturbative dynamics of scalar fields were studied, it was shown that noncommutativity can lead to  unfamiliar ultraviolet (UV) and the infrared (IR) behaviour    
of theories. In Refs.\cite{ym,arefeva} gauge theories were studied in the pure gauge theory whereas in Refs.\cite{hayakawa1,hayakawa2,sheikh2,krajewski,susskind} matter fields were introduced and detailed perturbative analysis of the UV and IR behaviour of noncommutative gauge theory were carried out to one loop. Due to the form of vertices in noncommutative gauge theory, certain phase factors appear in the Feynman integrals in loop diagrams. Depending on whether these phases contain the loop momenta or not, one speaks of nonplanar or planar diagrams, respectively.
While the Feynman integrals of planar diagrams in noncommutative gauge theory are exactly the same as in the commutative case, within  a constant factor, the nonplanar diagrams lead to important, rather unconventional effects in the short- and long-distance behaviour of the theory. 
\par
In this paper we study the Ward identities and calculate the U(1)-anomaly in noncommutative (NC)-U(1) gauge theory. Ward identities and anomalies are important for the renormalizability of the NC-gauge theories. Recent studies have shown that noncommutative QED shares many properties of the ordinary commutative non-Abelian gauge theories\cite{ym,arefeva}. As for certain gauge groups and their representations the anomaly cancels, it is an interesting question to know under what condition the anomaly cancels in NC-QED. 
\par\noindent
The study of the anomalies in noncommutative gauge theories is also intersesting in connection with the UV/IR mixing. As is well-known the U(1) anomaly is the result of a finite shift of integration variables in the {\it ultraviolet} (linearly) divergent Feynman integrals for triangle diagrams\cite{adler}. On the other hand the calculation of the commutative U(1)-anomaly in lattice gauge theory shows that a very similar shift of integration variables in the lattice regulated Feynman integrals for the triangle diagrams leads to a surface integral including an {\it infrared} singularity at the origin of the first Brillouine zone\cite{neda2}. There are other evidences which prove that in commutative gauge theory the axial anomaly seems to have its origin both in the high (UV) as well as in the low energy (IR) regime\cite{frishman}. It is therefore important to study the axial anomaly in detail in order to find its origin in noncommutative gauge theories. Anomalies arising from nonplanar integrals, where UV/IR mixing seems to appear, are studied in a second paper\cite{neda4} separately. 
\par 
In this paper, we have studied the axial anomaly arising from the planar diagrams of NC-U(1) gauge theory with fermions in the fundamental representation in $D=2,4$ dimensions.
The organization of the paper is  as follows. In Section 2 we introduce the noncommutative QED on {\bf R}$^D$. After listing  of Feynman rules of interest, gauge invariance of this theory is checked in a simple example of fermion-antifermion annihilation into two gauge bosons at the tree level (Section 3). As in the commutative non-Abelian gauge theories gauge invariance is guaranteed, by considering not only the diagrams of the  commutative QED but also a diagram containing the three-photon vertex.
\par
In Section 4, we then calculate the axial anomaly using  the method of point splitting for two- and four-dimensional NC-QED. 
Comparing to the commutative QED case, an additional diagram has to be considered to produce the non-linear term in the field strength tensor of NC-QED. Besides, from the start a non-trivial integration over noncommutative space-time coordinates must be performed in order to guarantee the $\star$-gauge invariance of the result.  
\par
In the commutative field theory, the axial anomaly can also be interpreted as a result of the non-invariance of the integration measure of the fermionic fields  under local axial gauge transformation. In noncommutative gauge theory, it is therefore interesting to check whether the usual path integral method of Fujikawa\cite{fuji} goes through. From the technical point of view the difference between this calculation and the standard one is that here, a $\star$-gauge {\it covariant} regularization leads to a $\star$-gauge {\it covariant} expression for the anomaly and as in the method of point-splitting gauge invariance can only be guaranteed after averaging over the densities. This has serious consequences for topological properties of the anomaly. Here, the result for the axial anomaly seems to be the same as in the commutative QED with the replacement of the ordinary product of functions in the coordinate space with the $\star$-product. This replacement, however, is not {\it a priori}. As for the case of instantons in the noncommutative space an {\it a priori} modification of the usual product with the $\star$-product of function does not work\cite{nekrasov}. It is therefore important to carry out the calculation explicitly in NC-U(1) gauge theory. 
\par
We then turn to the triangle diagrams which, reproducing the previous result after inclusion of higher loop corrections.  As it turns out the triangle diagarms are all planar. This means that the corresponding Feynman integrals differ from the usual commutative triangle integrals in an overall phase factor, which only depends on the external loop momenta. These planar phase can be brought out of the integrals and the rest of the calculation is just the standard calculation of the anomaly in the commutative QED.\footnote{In Ref.\cite{neda4} we study the effects of {\it nonplanar diagrams} on global U(1) anomaly.} The expression for the axial anomaly turns out to be $\star$-gauge invariant only after an integration over noncommutative space-time coordinates, which then gives the Chern class. This calculation which yields a naively predictable result is performed here explicitly in order to make a connection to another interesting method, where by using the current algebra of NC-QED, a separation between a ''formal'' part and a potentially ''anomalous'' part of the divergence of the vertex function can be made, which is necessary in the calculation of the axial anomaly for other noncommutative gauge theories, e.g. NC-$U\left(N\right)$\cite{neda4}. Here, as a by product, the current algebra of NC-U(1) gauge theory is presented. The non-trivial expressions for the equal-time commutation relations of two-currents contain the novel effect of being the difference of these currents at different space-time points, which cannot be obtained from the Abelian or even non-Abelian commutative equal-time commutation relations by just replacing the ordinary product with the $\star$-product.  
Section 5 is devoted to discussions.
\par\vskip0.5cm\par
\section{{Noncommutative Gauge Theory (NCGT)}}
\setcounter{section}{2}
\setcounter{equation}{0}
\noindent
In the noncommutative geometry formulation, geometric spaces are described by a $C^{\star}$-algebra, which is in general  not commutative. 
The familiar product of functions is therefore replaced with the $\star$-{\it product}, which is characteristic for the noncommutative geometry and is given by:
\begin{eqnarray}\label{F21}  
f\left(x\right)\star g\left(x\right)\equiv e^{\frac{i\theta_{\mu\nu}}{2}\ \frac{\partial}{\partial\xi_{\mu}}\ \frac{\partial}{\partial\zeta_{\nu}} }f\left(x+\xi\right)g\left(x+\zeta\right)\bigg|_{\xi=\zeta=0},
\end{eqnarray}
where $\theta_{\mu\nu}$ is a real constant antisymmetric back\-ground, and re\-flects the noncom\-mu\-ta\-ti\-vi\-ty of the coordinates of {\bf R}$^D$\cite{witten}:
\begin{eqnarray}\label{F22}
[x_{\mu},x_{\nu}]=i\theta_{\mu\nu},
\end{eqnarray}
The usual algebra of functions in {\bf R}$^D$ is therefore modified to a noncommutative associative algebra such that $f\star g=fg+\frac{1}{2}i\theta^{\mu\nu}\partial_{\mu}f\partial_{\nu}g+{\cal{O}}\left(\theta^{2}\right)$, with  coefficients of each power being local differential expressions bilinear in $f$ and $g$. The $\star$-product satisfies the identity:
\begin{eqnarray}
\int\limits_{-\infty}^{+\infty}d^{D}x\ f\left(x\right)\star g\left(x\right) =
\int\limits_{-\infty}^{+\infty}d^{D}x\ g\left(x\right)\star f\left(x\right)=\int\limits_{-\infty}^{+\infty}d^{D}x\ f\left(x\right)g\left(x\right),\label{F23a}
\end{eqnarray}
where the product on the last expression on the right hand side (r.h.s) of this equation is the usual product of  both functions $f\left(x\right)$ and $g\left(x\right)$. The associativity leads further to:
\begin{eqnarray}\label{F23b}
\int\limits_{-\infty}^{+\infty}d^{D}x\ \left(f_{1}\star f_{2}\star f_{3}\right)\left(x\right) =
\int\limits_{-\infty}^{+\infty}d^{D}x\ \left(f_{3}\star f_{1}\star f_{2}\right)\left(x\right) = \int\limits_{-\infty}^{+\infty}d^{D}x\ \left(f_{2}\star f_{3}\star f_{1}\right)\left(x\right).
\end{eqnarray}
\par\vskip0.3cm\par\noindent
{\bf Classical Action}
\par\vskip0.3cm\par\noindent
The noncommutative Yang-Mills action is introduced in Refs.\cite{ym,arefeva}, and the coupling of the U(1) gauge fields with the matter fields is discussed in\cite{hayakawa1,hayakawa2}. The noncommutative U(1) gauge connection is given by\cite{connes,arefeva}:
\begin{eqnarray}\label{F25}
D_{\mu}=\partial_{\mu}+ig[A_{\mu},\ \  ]_{\star},
\end{eqnarray}
where $[f_{1},f_{2}]_{\star}$ is the Moyal bracket defined by:
\begin{eqnarray}\label{F26}
[f_{1},f_{2}]_{\star}=f_{1}\star f_{2}-f_{2}\star f_{1}.
\end{eqnarray}
This leads to the definition of the field strength tensor 
\par\vskip0.5cm
\hspace{6cm}{$[D_{\mu},D_{\nu}]_{\star}=igF_{\mu\nu}\left(x\right)$},
\par\noindent
with 
\begin{eqnarray}\label{F27}
F_{\mu\nu}\left(x\right)&\equiv& \partial_{\mu}A_{\nu}\left(x\right)-\partial_{\nu}A_{\mu}\left(x\right)+ig\big[A_{\mu}\left(x\right),A_{\nu}\left(x\right)\big]_{\star}.
\end{eqnarray}
The pure noncommutative U(1) gauge action is then given by:
\begin{eqnarray}\label{F28}
S_{G}[A_{\mu}]= -\frac{1}{4}\int d^{D}x \ F_{\mu\nu}\left(x\right)\star F^{\mu\nu}\left(x\right).
\end{eqnarray}  
Under an arbitrary local gauge transformation the gauge field $A_{\mu}\left(x\right)$ transforms as\cite{hayakawa1,hayakawa2}:
\begin{eqnarray}\label{F29}
A_{\mu}\left(x\right)\to A'_{\mu}\left(x\right)=U\left(x\right)\star A_{\mu}\left(x\right)\star U^{-1}\left(x\right)+\frac{i}{g}\big[\partial_{\mu}U\left(x\right)\big]\star U^{-1}\left(x\right),
\end{eqnarray}   
where
\begin{eqnarray}\label{F210}
U\left(x\right)&\equiv& \left(e^{ig\alpha\left(x\right)}\right)_{\star}=1+ig\alpha\left(x\right)-\frac{g^{2}}{2!}\alpha\left(x\right)\star\alpha\left(x\right)+{\cal{O}}\left(\alpha_{\star}^{3}\right).
\end{eqnarray}
The subscript $\star$ denotes $\star$-multiplication in the products. Similarly $U^{-1}\left(x\right)\equiv \left(e^{-ig\alpha\left(x\right)}\right)_{\star}$, which can be shown to be the inverse of the $U\left(x\right)$, i.e. $U\left(x\right)\star U^{-1}\left(x\right)=1$. Furthermore, the covariant derivative transforms as:
\begin{eqnarray}\label{F211}
D_{\mu}\to U\left(x\right)\star D_{\mu}\star U^{-1}\left(x\right). 
\end{eqnarray}
For infinitesimal gauge transformation parameter $\alpha\left(x\right)$ the transformation law for the gauge field and the field strength tensor is given by\cite{witten}:
\begin{nedalph}
A_{\mu}'\left(x\right)&=&A_{\mu}\left(x\right)-\partial_{\mu}\alpha\left(x\right)+ig[\alpha\left(x\right),A_{\mu}\left(x\right)]_{\star} .\label{F212a}\\
F'_{\mu\nu}\left(x\right)&=&F_{\mu\nu}\left(x\right)+ig[\alpha\left(x\right),F_{\mu\nu}\left(x\right)]_{\star}.\label{F212b}
\end{nedalph}
The above gauge action [Eq. (\ref{F28})] is then invariant under infinitesimal local gauge transformations (\ref{F212b}).  
Note that the expression $F_{\mu\nu}\star F^{\mu\nu}$ is not $\star$-gauge invariant by itself, in contrast to the local trace of the corresponding term in non-Abelian commutative gauge theories.
The $\star$-gauge invariance of the gauge action (\ref{F28}) under gauge transformation (\ref{F29}) is only guaranteed after the integration of this expression over the noncommutative space-time coordinates $x$.   
\par
The matter fields are similarly introduced\cite{hayakawa1}:
\begin{eqnarray}\label{F214}
S_{F}[\overline{\psi},\psi]= \int d^{D}x \bigg[i\overline{\psi}\left(x\right)\gamma^{\mu}\star D_{\mu}\psi\left(x\right)-m\overline{\psi}\left(x\right)\star\psi\left(x\right)\bigg],
\end{eqnarray}
where the covariant derivative is defined by:
\begin{eqnarray}\label{F215}
D_{\mu}\psi\left(x\right)\equiv\partial_{\mu}\psi\left(x\right)+igA_{\mu}\left(x\right)\star\psi\left(x\right).
\end{eqnarray}
The action (\ref{F214}) is invariant under the local  gauge transformations of the gauge fields (\ref{F29}) and the local gauge transformations of the matter fields:
\begin{eqnarray}\label{F216}
\psi\left(x\right)\to\psi'\left(x\right)=U\left(x\right)\star \psi\left(x\right),\hspace{1cm}\overline{\psi}\left(x\right)\to \overline{\psi}'\left(x\right)=\overline{\psi}\left(x\right)\star U^{-1}\left(x\right).
\end{eqnarray}
Hence for infinitesimal local gauge transformations we have:
\begin{eqnarray}\label{F217}
\overline{\psi}'\left(x\right)\approx\overline{\psi}\left(x\right)-ig\overline{\psi}\left(x\right)\star\alpha\left(x\right)\hspace{0.5cm}\mbox{and}\hspace{0.5cm}
\psi'\left(x\right)\approx\psi\left(x\right)+ig\alpha\left(x\right)\star \psi\left(x\right).
\end{eqnarray}
The total action in the noncommutative U(1) gauge theory therefore reads: 
\begin{eqnarray}\label{F218}
S_{tot.}[A_{\mu},\overline{\psi},\psi]=
-\frac{1}{4}\int d^{D}x \ F_{\mu\nu}\left(x\right)\star F^{\mu\nu}\left(x\right)+ \int d^{D}x \ \overline{\psi}\left(x\right)\star
\left(i\DS-m\right)\psi\left(x\right).
\end{eqnarray}
\par\vskip0.5cm\par\noindent
\section{Feynman Rules and Ward-Identity of NC-QED}
\setcounter{section}{3}
\setcounter{equation}{0}
\noindent
Here, in analogy to commutative non-Abelian gauge theories, gauge fixing is necessary and leads to noncommutative Faddeev-Popov-ghosts. The Feynman rules for gauge fields, matter fields and ghosts are given in  Ref.\cite{hayakawa2}. We will list here only the rules we will need in this paper: 
\par
The fermion  and photon propagators do not change in NC-QED. For the fer\-mion\--pro\-pa\-ga\-tor we have:
\begin{nedalph}\label{FX1a}
\SetScale{0.8}
    \begin{picture}(80,20)(0,0)
    \Vertex(0,0){2}
    \Line(0,0)(50,0)
    \Vertex(50,0){2}
    \Text(20,-10)[]{$p$}
    \end{picture}
\hspace{1cm} S\left(p\right)=\frac{i}{\PS-m},
\end{eqnarray}
whereas the photon propagator in Feynman gauge reads:
\begin{eqnarray}\label{FX1b}
\SetScale{0.8}
    \begin{picture}(80,20)(0,0)
    \Vertex(0,0){2}
    \Photon(0,0)(50,0){3}{6}
    \Vertex(50,0){2}
    \Text(20,-10)[]{$k$}
    \Text(0,-10)[]{$\mu$}
    \Text(40,-10)[]{$\nu$}
    \end{picture}
\hspace{1cm}
D_{\mu\nu}\left(k\right)=\frac{-ig_{\mu\nu}}{k^{2}}.
\end{eqnarray}
The vertex of two fermions and one gauge field is given by:
\vskip0.2cm
\begin{eqnarray}\label{FX1c}
\SetScale{0.8}
    \begin{picture}(50,20)(0,0)
    \Vertex(0,0){2}
	\ArrowArc(0,0)(20,250,300)
    \Photon(0,0)(0,20){2}{4}
    \LongArrow(5,18)(5,12)
    \ArrowLine(-20,-20)(0,0)
    \ArrowLine(20,-20)(0,0)
    \Text(20,15)[]{$k_{1},\mu$}
    \Text(-25,-15)[]{$p_{1}$}
    \Text(25,-15)[]{$p_{2}$}
    \end{picture}
\hspace{0.5cm}
V_{\mu}\left(p_{1},p_{2};k_{1}\right)=ig\left(2\pi\right)^{4}\delta^{4}\left(p_{1}+p_{2}+k_{1}\right)\gamma_{\mu}\ \mbox{exp}\left(\frac{-i\theta_{\eta\sigma}}{2}p_{1}^{\eta}p_{2}^{\sigma}\right).
\end{eqnarray}
\vskip0.5cm
The last term in the definition of field strength tensor containing the Moyal bracket, leads to three- and four-gauge vertices in the NC-QED. The three-photon vertex is:
\vskip0.2cm
\begin{eqnarray}\label{FX1d}
\SetScale{0.8}
    \begin{picture}(50,20)(0,0)
    \Vertex(0,0){2}
    \Photon(0,0)(0,20){2}{4}
    \LongArrow(-10,20)(-10,12)
    \LongArrow(-25,-15)(-20,-8)
    \LongArrow(25,-15)(20,-8)
    \Photon(-20,-20)(0,0){2}{4}
    \Photon(20,-20)(0,0){2}{4}
    \Text(20,15)[]{$k_{1},\mu_{1}$}
    \Text(-35,-15)[]{$k_{2},\mu_{2}$}
    \Text(35,-15)[]{$k_{3},\mu_{3}$}
    \end{picture}
\hspace{0.2cm}
\lefteqn{W_{\mu_{1}\mu_{2}\mu_{3}}\left(k_{1},k_{2},k_{3}\right)=-2g\left(2\pi\right)^{4}\delta^{4}\left(k_{1}+k_{2}+k_{3}\right)   \sin\left(\frac{i\theta_{\eta\sigma}}{2}k_{1}^{\eta}k_{2}^{\sigma}\right) } \nonumber\\
&&\hspace{0.cm}\times
\bigg[
g_{\mu_{1}\mu_{2}}\left(k_{1}-k_{2}\right)_{\mu_{3}} +
g_{\mu_{1}\mu_{3}}\left(k_{3}-k_{1}\right)_{\mu_{2}}+
g_{\mu_{3}\mu_{2}}\left(k_{2}-k_{3}\right)_{\mu_{1}}  
\bigg] .
\end{nedalph}
In analogy to the non-Abelian field theory all vertices involve the same coupling constant. 
\par\vskip0.3cm\par\noindent
{\bf Check of the Ward-Identity in a simple case}
\par\vskip0.3cm\par\noindent
To see a parallel between NC-QED and commutative non-Abelian gauge theory we study the Ward identity at tree level. In this and the rest of the paper we follow Refs. \cite{peskin,weinberg} closely. As in the commutative gauge theory the amplitude of gauge boson production is expected to obey the Ward identity: 
\vskip0.5cm
\begin{eqnarray}\label{FY}
k^{\mu}\left(
\SetScale{0.8}
    \begin{picture}(80,30)(0,0)
    \Photon(0,0)(50,0){3}{6}
    \GOval(60,0)(10,10)(0){0.8}
    \Photon(65,10)(80,30){2}{4}
    \Photon(65,-10)(80,-30){2}{4}
    \Line(70,5)(85,10)
    \Line(70,-5)(85,-10)
    \end{picture} 
\right)=0.
\end{eqnarray}
\vskip0.5cm
Let us check this identity in the simple case of the lowest order diagrams contributing to fermion-antifermion annihilation into a pair of photons. In order $g^{2}$ there are three diagrams, shown in Fig.  [1]. The first two diagrams are similar to QED diagrams. They sum to:
\begin{eqnarray}\label{FX2}
\lefteqn{\hspace{-1cm}i{\cal{M}}_{a+b}^{\mu\nu}\epsilon_{\mu}^{*}\left(k_{1}\right)\epsilon_{\nu}^{*}\left(k_{2}\right)
=\left(ig\right)^{2}\overline{v}\left(p_{+}\right)e^{-\frac{i\theta_{\eta\sigma}}{2}p^{\eta}p_{+}^{\sigma}}\Bigg[\gamma^{\mu}\frac{i}{\PS-\KS_{2}-m}\gamma^{\nu}e^{-\frac{i\theta_{\eta\sigma}}{2}\left(p+p_{+}\right)^{\eta}k_{2}^{\sigma}}  }\nonumber\\
&&
+\gamma^{\nu}\frac{i}{\KS_{2}-\PS_{+}-m}\gamma^{\mu}e^{+\frac{i\theta_{\eta\sigma}}{2}\left(p+p_{+}\right)^{\eta}k_{2}^{\sigma}}  \Bigg]u\left(p\right)\epsilon_{\mu}^{*}\left(k_{1}\right)\epsilon_{\nu}^{*}\left(k_{2}\right).
\end{eqnarray}
Here ${\cal{M}}_{a+b}^{\mu\nu}$ denotes the contribution of both diagrams [1a] and [1b]; $\epsilon\left(k_{i}\right)$'s are photon polarization vectors. For physical polarization they satisfy $k_{i}^{\mu}\epsilon_{\mu}\left(k_{i}\right)=0$. Note that the exponential functions containing the noncommutativity parameter $\theta_{\eta\sigma}$ are obtained by making use of the definition of the usual vertex from the Eq. (\ref{FX1c}).
\par
To check the Ward identity we replace $\epsilon_{\nu}^{*}\left(k_{2}\right)$ in the Eq. (\ref{FX2}) by $k_{2\nu}$. This leads to:
\begin{eqnarray}\label{FX3}
\lefteqn{\hspace{-1cm}i{\cal{M}}_{a+b}^{\mu\nu}\epsilon_{\mu}^{*}\left(k_{1}\right)k_{2\nu}
=\left(ig\right)^{2}\overline{v}\left(p_{+}\right)e^{-\frac{i\theta_{\eta\sigma}}{2}p^{\eta}p_{+}^{\sigma}}\Bigg[\gamma^{\mu}\frac{i}{\PS-\KS_{2}-m}\ \KS_{2}\ e^{-\frac{i\theta_{\eta\sigma}}{2}\left(p+p_{+}\right)^{\eta}k_{2}^{\sigma}}  }\nonumber\\
&&
+\KS_{2}\ \frac{i}{\KS_{2}-\PS_{+}-m}\gamma^{\mu}e^{+\frac{i\theta_{\eta\sigma}}{2}\left(p+p_{+}\right)^{\eta}k_{2}^{\sigma}}  \Bigg]u\left(p\right)\epsilon_{\mu}^{*}\left(k_{1}\right).
\end{eqnarray}
Then use of equations of motion leads to: 
\begin{eqnarray}\label{FX5}
i{\cal{M}}_{a+b}^{\mu\nu}\epsilon_{\mu}^{*}\left(k_{1}\right)k_{2\nu}=2g^{2} 
\overline{v}\left(p_{+}\right)\gamma^{\mu}u\left(p\right)\epsilon_{\mu}^{*}\left(k_{1}\right)
e^{-\frac{i\theta_{\eta\sigma}}{2}p^{\eta}p_{+}^{\sigma}}
\sin\left(\frac{\theta_{\eta\sigma}}{2}\left(p+p_{+}\right)^{\eta}k_{2}^{\sigma}\right)
.
\end{eqnarray}
Notice that the above sine-function, which contains the parameter $\theta_{\eta\sigma}$, replaces the structure constant $f_{abc}$ in the non-Abelian gauge theory. 
\par
The contribution of the diagram [1c], which in the non-Abelian commutative case is needed to satisfy the Ward identity (\ref{FY}), is given by:
\begin{eqnarray}\label{FX6}
\lefteqn{\hspace{-1cm}i{\cal{M}}_{c}^{\mu\nu}\epsilon_{\mu}^{*}\left(k_{1}\right)\epsilon_{\nu}^{*}\left(k_{2}\right)=+2ig^{2}e^{-\frac{i\theta_{\eta\sigma}}{2}p^{\eta}p_{+}^{\sigma}}\sin\left(\frac{\theta_{\eta\sigma}}{2}\left(p+p_{+}\right)^{\eta}k_{2}^{\sigma}\right)\overline{v}\left(p_{+}\right)\gamma_{\rho}u\left(p\right)\left(\frac{-i}{k_{3}^{2}}\right) 
}\nonumber\\
&&\times\bigg[
g^{\mu\nu}\left(k_{2}-k_{1}\right)^{\rho}+
g^{\nu\rho}\left(k_{3}-k_{2}\right)^{\mu}+
g^{\mu\rho}\left(k_{1}-k_{3}\right)^{\nu}
\bigg] \epsilon_{\mu}^{*}\left(k_{1}\right)\epsilon_{\nu}^{*}\left(k_{2}\right),
\end{eqnarray}
where $k_{3}=-k_{1}-k_{2}$ due to energy-momentum conservation. Replacing $\epsilon_{\nu}\left(k_{2}\right)$ with $k_{2\nu}$ and using the momentum conservation the expression in the brackets reads:
\begin{eqnarray}\label{FX7}
\lefteqn{\hspace{-3cm}k_{2\nu}\bigg[
g^{\mu\nu}\left(k_{2}-k_{1}\right)^{\rho}+
g^{\nu\rho}\left(k_{3}-k_{2}\right)^{\mu}+
g^{\mu\rho}\left(k_{1}-k_{3}\right)^{\nu}
\bigg]=}\nonumber\\
&&\hspace{2cm}=
k_{3}^{2}g^{\mu\rho}-k^{3\mu}k^{3\rho}-k_{1}^{2}g^{\mu\rho}+k^{1\mu}k^{1\rho}.
\end{eqnarray}
Taking the gauge boson  with momentum $k_{1}$ to be on-shell and with transverse polarization, we have $k_{1}^{2}=0$ \footnote{Contrast this with the dispersion relation of Ref.\cite{susskind}. That dispersion relation is a consequence of the adjoint representation of matter fields.} as well as $k_{1}^{\mu}\epsilon_{\mu}\left(k_{1}\right)=0$. Then the third and the fourth terms on the r.h.s of the above equation vanish.  Furthermore, the term $k^{3\mu}k^{3\rho}$ vanishes when it is contracted with the fermion current. In the remaining term, the factor $k_{3}^{2}$ cancels the photon propagator, and we are left with
\begin{eqnarray}\label{FX8}
i{\cal{M}}_{c}^{\mu\nu}\epsilon_{\mu}^{*}\left(k_{1}\right)k_{2\nu}=-2g^{2}\overline{v}\left(p_{+}\right)\gamma_{\mu}u\left(p\right)\epsilon_{\mu}^{*}\left(k_{1}\right)e^{-\frac{i\theta_{\eta\sigma}}{2}p^{\eta}p_{+}^{\sigma}}\sin\left(\frac{\theta_{\eta\sigma}}{2}\left(p+p_{+}\right)^{\eta}k_{2}^{\sigma}\right),
\end{eqnarray}
thus cancelling (\ref{FX5}). 
The equality of the coupling constant in the fermion-photon and three-photon vertices guarantees the validity of the Ward identity as in the commutative non-Abelian gauge theories. 
\par\vskip0.5cm\noindent
\section{U(1)-Anomaly in NC-QED}
\par\vskip0.3cm\par\noindent
{\bf Point-Splitting Method}\label{S41}
\par\vskip0.3cm\par\noindent
\noindent{\it i) U(1)-Anomaly in Two Dimensions}
\setcounter{section}{4}
\setcounter{equation}{0}
\par\noindent
Before we take up the four dimensional anomaly, let us calculate the anomaly for two dimensional NC-QED, with matter fields Lagrangian:
\begin{eqnarray}\label{F31}
{\cal{L}}=i\overline{\psi}\left(x\right)\star\gamma^{\mu}D_{\mu}\psi\left(x\right), 
\end{eqnarray}
which leads to the equation of motion
\begin{eqnarray}\label{F32}
\gamma^{\mu}\partial_{\mu}\psi\left(x\right)=-ig\gamma^{\mu}A_{\mu}\left(x\right)\star\psi\left(x\right),\hspace{0.5cm}\mbox{and}\hspace{0.5cm}\partial_{\mu}\overline{\psi}\left(x\right)\gamma^{\mu}=ig\overline{\psi}\left(x\right)\star A_{\mu}\left(x\right)\gamma^{\mu}.
\end{eqnarray}
The anomaly to calculate is that of the axial vector current:\footnote{As we have shown in Ref.\cite{neda4} U(1) gauge theory with fundamental matter field coupling has three different currents, which all, expressing the same global symmetry of the noncommutative action, lead to the same classically conserved charge. Here we intend to work only with one of these currents.}
\begin{eqnarray}\label{F33}
J^{\mu\left(5\right)}\left(x\right)\equiv i\overline{\psi}\left(x\right)
\gamma^{\mu}\gamma^{5}\star\psi\left(x\right).
\end{eqnarray}
This product of local operators is singular. The current is regularized by placing  the two fermion fields at different ''lattice''  points separated by a distance $\epsilon$:
\begin{eqnarray*}
J^{\mu\left(5\right)}\equiv i\overline{\psi}\left(x\right)\gamma^{\mu}\gamma^{5}\star\psi\left(x\right)\longrightarrow i\overline{\psi}\left(x+\epsilon/2\right)\gamma^{\mu}\gamma^{5}\star\psi\left(x-\epsilon/2\right),
\end{eqnarray*}
which is not, however, invariant under local gauge transformations defined in the Eq. (\ref{F216}). To make it $\star$-gauge invariant, introduce the link variable ${\cal{U}}\left(y,x\right)$:
\begin{eqnarray}\label{F34}
{\cal{U}}\left(y,x\right)\equiv \exp\bigg[-ig\int\limits_{x}^{y} dz^{\mu}  A_{\mu}\left(z\right)\bigg],
\end{eqnarray}  
which transforms under the local gauge transformation as:
\begin{eqnarray}\label{F35}
{\cal{U}}'\left(y,x\right)=U\left(y\right)\star {\cal{U}}\left(y,x\right)\star U^{-1}\left(x\right),
\end{eqnarray}
where $U\left(x\right)$ is the unitary matrix defined in the Eq. (\ref{F210}) and $U\left(x\right)\star U^{-1}\left(x\right)=1$. The axial vector current in its point splitted, $\star$-gauge invariant version is then given by:
\begin{eqnarray}\label{F36}
J^{\mu\left(5\right)}\left(x\right)=\lim\limits_{\epsilon\to 0}\bigg\{\overline{\psi}\left(x+\epsilon/2\right)\gamma^{\mu}\gamma^{5}\star{\cal{U}}\left(x+\epsilon/2,x-\epsilon/2\right)\star\psi\left(x-\epsilon/2\right)\bigg\}.
\end{eqnarray} 
Keeping terms up to the first order in $\epsilon$ in ${\cal{U}}\left(y_{+},y_{-}\right)$ for $y_{\pm}=x\pm\epsilon/2$ we get:
\begin{eqnarray}\label{F37}
{\cal{U}}\left(x+\epsilon/2,x-\epsilon/2\right) = 1-ig\epsilon^{\nu}A_{\nu}\left(x\right)+{\cal{O}}\left(\epsilon^{2}\right), 
\end{eqnarray}  
and
\begin{eqnarray}\label{F38}
\partial_{\mu}{\cal{U}}\left(x+\epsilon/2,x-\epsilon/2\right)=-ig\epsilon^{\nu}\partial_{\mu}A_{\nu}\left(x\right)+{\cal{O}}\left(\epsilon^{2}\right).
\end{eqnarray}
Then we use the equations of motion (\ref{F32}) and the expression (\ref{F38}) in the Eq. (\ref{F36}) to obtain:
\begin{eqnarray}\label{F310}
\lefteqn{
\partial_{\mu}J^{\mu\left(5\right)}\left(x\right)=\lim\limits_{\epsilon\to 0}\  \overline{\psi}\left(x+\epsilon/2\right) \gamma^{\mu}\gamma^{5}\star\bigg[ig\{A_{\mu}\left(x+\epsilon/2\right)-A_{\mu}\left(x-\epsilon/2\right)\}
}\nonumber\\
&&+g^{2}\epsilon^{\nu}\ \{A_{\mu}\left(x+\epsilon/2\right)\star A_{\nu}\left(x\right)-A_{\nu}\left(x\right)\star A_{\mu}\left(x-\epsilon/2\right)\}-ig\epsilon^{\nu}\partial_{\mu}A_{\nu}\left(x\right)\bigg]\star\psi\left(x-\epsilon/2\right).\nonumber\\
\end{eqnarray}  
Note that in the standard  derivation of the axial anomaly for two-dimensional commutative QED the first two terms on the last line of the above equation cancel due to the {\sl commutativity} of the ordinary product. Here, exactly this term leads to the additional term $[A_{\mu}, A_{\nu}]_{\star}$ in the definition of  field strength tensor  in NC-QED [see Eq. (\ref{F27})]. 
\par
Expanding the expression on the r.h.s. of Eq. (\ref{F310}) and keeping  the terms of order $\epsilon^{\nu}$, the divergence of the axial current reads:
\begin{eqnarray}\label{F311}
\partial_{\mu}J^{\mu\left(5\right)}\left(x\right)=\lim\limits_{\epsilon\to 0}\bigg[-ig\ \epsilon^{\nu}\  \overline{\psi}\left(x+\epsilon/2\right)\gamma^{\mu}\gamma^{5}\star F_{\mu\nu}\left(x\right)\star\psi\left(x-\epsilon/2\right)\bigg],
\end{eqnarray} 
where $F_{\mu\nu}\left(x\right)$ is defined in the Eq. (\ref{F27}). 
Let us take the expectation value of the above equation in the presence of a fixed background gauge field. To obtain a $\star$-gauge invariant result, we take the integral over  two dimensonal space-time coordinates of this equation. Only after this integration one can make use of Eq. (\ref{F23b}) to find:
\begin{eqnarray}\label{F312}
\lefteqn{\hspace{-2cm}
\int \limits_{-\infty}^{+\infty}d^{2}x\ \bigg<\partial_{\mu}J^{\mu\left(5\right)}\left(x\right)\bigg>
=
  \lim\limits_{\epsilon\to 0}\bigg[-ig\ \epsilon^{\nu}\  \int\limits_{-\infty}^{+\infty}d^{2}x\ 
F_{\mu\nu}\left(x\right)\star \bigg<
\psi\left(x-\epsilon/2\right)\gamma^{\mu}\gamma^{5} \star\overline{\psi}\left(x+\epsilon/2\right)\bigg>\bigg]
}\nonumber\\
&=&
-ig\ \int\limits_{-\infty}^{+\infty}d^{2}x\ 
F_{\mu\nu}\left(x\right)\ \lim\limits_{\epsilon\to 0}\epsilon^{\nu}\  
\bigg<\psi\left(x-\epsilon/2\right)\gamma^{\mu}\gamma^{5} \star\overline{\psi}\left(x+\epsilon/2\right)\bigg>,
\end{eqnarray}
where the $\star$-product between $F_{\mu\nu}\left(x\right)$ and the rest of the expression is removed by using the Eq. (\ref{F23a}). 
In the momentum space the vacuum expectation value of the $\star$-product of $\psi$ and $\overline{\psi}$ is given by:  
\begin{eqnarray*}
{\cal{I}}\equiv\bigg<\psi\left(y\right)\gamma^{\mu}\gamma^{5}\star\overline{\psi}\left(z\right)\bigg>&=& \bigg<\int\limits_{-\infty}^{+\infty} \frac{d^{2}p}{\left(2\pi\right)^{2}}\  \frac{d^{2}q}{\left(2\pi\right)^{2}}\ \tilde{\psi}\left(p\right)\ \gamma^{\mu}\gamma^{5}\  \tilde{\overline{\psi}}\left(q\right)\ e^{-ipy}\star e^{+iqz}\bigg>.
\end{eqnarray*}
Here $y=x-\frac{\epsilon}{2}$ and $z=x+\frac{\epsilon}{2}$. 
In the zeroth order of perturbative expansion\footnote{Higher order terms are less divergent and vanish after contracting with $\epsilon^{\nu}$ and taking the limit $\epsilon\to 0$.} the contraction of fermion-antifermion leads to a term proportional to $\delta^{\left(2\right)}\left(p-q\right)\Delta\left(p\right)$, where $\Delta\left(p\right)=\frac{i\PS}{p^{2}}$ is the fermion propagator in massless case. The phase factor exp$\left(\frac{i}{2}\theta_{\eta\sigma}p^{\eta}q^{\sigma}\right)$, which appears in the $\star$-product of the exponential functions in the above expression vanishes due to the $\delta$-function. We arrive at:
\begin{eqnarray}\label{F313}
{\cal{I}}^{\left(0\right)}&=& -\int\limits_{-\infty}^{+\infty} \frac{d^{2}p}{\left(2\pi\right)^{2}} \ \mbox{Tr}\left(\gamma^{\alpha}\gamma^{\mu}\gamma^{5}\right)\   \frac{ip_{\alpha}\ e^{ip\epsilon} }{p^{2}}=-\mbox{Tr}\left(\gamma^{\alpha}\gamma^{\mu}\gamma^{5}\right)\ \frac{\partial}{\partial\epsilon_{\alpha}}\int\limits_{-\infty}^{+\infty} \frac{d^{2}p}{\left(2\pi\right)^{2}}\ \frac{e^{ip\epsilon}}{p^{2}}\nonumber\\
&=&-\frac{i}{4\pi}\mbox{Tr}\left(\gamma^{\alpha}\gamma^{\mu}\gamma^{5}\right)\frac{\partial}{\partial \epsilon_{\alpha}}\mbox{log}\epsilon^{2}=-\frac{i}{2\pi}\mbox{Tr}\left(\gamma^{\alpha}\gamma^{\mu}\gamma^{5}\right)   
\frac{\epsilon_{\alpha}}{\epsilon^{2}},
\end{eqnarray}
and therefore:
\begin{eqnarray*}
{\cal{I}}\left(\epsilon\right)\equiv\bigg<\psi\left(x-\epsilon/2\right)\ \gamma^{\mu}\gamma^{5}\ \star\overline{\psi}\left(x+\epsilon/2\right)\bigg>=-\frac{i}{2\pi}\mbox{Tr}\left(\gamma^{\alpha}\gamma^{\mu}\gamma^{5}\right)   
\frac{\epsilon_{\alpha}}{\epsilon^{2}}+{\cal{O}}\left(\log{\epsilon}\right).
\end{eqnarray*}
As we can see the contraction of the fermion-antifermion  fields are singular for $\epsilon\to 0$. A finite result can nevertheless be obtained after putting this result in the Eq. (\ref{F312}) and taking the limit over $\epsilon\to 0$:
\begin{eqnarray}\label{F316}
\int \limits_{-\infty}^{+\infty}d^{2}x\ \bigg<\partial_{\mu}J^{\mu\left(5\right)}\left(x\right)\bigg>&=&
-ig\int\limits_{-\infty}^{+\infty}d^{2}x\ F_{\mu\nu}\left(x\right)\lim\limits_{\epsilon\to 0}\ \epsilon^{\nu}\ {\cal{I}}\left(\epsilon\right)\nonumber\\
&=& -\frac{g}{2\pi}\mbox{Tr}\left(\gamma^{\alpha}\gamma^{\mu}\gamma^{5}\right)\int\limits_{-\infty}^{+\infty}F_{\mu\nu}\left(x\right)\ \lim\limits_{\epsilon\to 0}\ \frac{\epsilon_{\alpha}\epsilon^{\nu}}{\epsilon^{2}}.
\end{eqnarray}
Using equation:
\begin{eqnarray}\label{F315}
\lim\limits_{\epsilon\to 0}\frac{\epsilon_{\alpha}\epsilon^{\nu}}{\epsilon^{2}}=\frac{1}{D}\delta^{\nu}_{\alpha}, 
\end{eqnarray}
in $D=2$ dimensions, and the relation Tr$\left(\gamma^{\alpha}\gamma^{\mu}\gamma^{5}\right)=2\varepsilon^{\alpha\mu}$, we obtain the axial anomaly in two-dimensional massless NC-QED:
\begin{eqnarray}\label{X1}
\int \limits_{-\infty}^{+\infty}d^{2}x\ \big<\partial_{\mu}J^{\mu\left(5\right)}\left(x\right)\big>= \frac{g}{2\pi}\ \varepsilon^{\mu\nu}\int\limits_{-\infty}^{+\infty}d^{2}x \ F_{\mu\nu}\left(x\right). 
\end{eqnarray}
At this stage two comments are in order: first the anomaly expression (\ref{X1}) is gauge invariant because of the space and time integrals. Secondly, we have allowed for the noncommutativity of space and time coordinates in two dimensions.
\par\vskip1cm
\noindent{\it ii) U(1)-Anomaly in Four Dimensions}
\par\noindent
We will now consider the four dimensional analogue of the above two-dimensional massless NC-QED. By taking the divergence of the axial vector current (\ref{F33}) all of the manipulations leading to Eq. (\ref{F312}) still go through.    
\par
The vacuum expectation value:
\begin{eqnarray}\label{F335e}
{\cal{T}}\left(x,\epsilon\right)=\bigg<\psi\left(x-\epsilon/2\right)\gamma^{\mu}\gamma^{5}\star\overline{\psi}\left(x+\epsilon/2\right)\bigg>,
\end{eqnarray}
for the four dimensional case is slightly differently calculated. 
Here, a perturbative expansion in a non-zero background gauge field with the interaction
\begin{eqnarray}\label{F336e}
S_{I}\equiv -g\int\limits_{-\infty}^{+\infty}d^{4}y\ \left(\overline{\psi}\star \AS\star\psi\right)\left(y\right) = -g\int\limits_{-\infty}^{+\infty}d^{4}y\ \left(\psi\  \gamma^{\beta}\star\overline{\psi}\right)\left(y\right)\ A_{\beta}\left(y\right),
\end{eqnarray}
is necessary. In a perturbation series the contribution of the n-th order receives a factor:
\begin{eqnarray}\label{F337e}
\frac{\left(+ig\right)^{n}}{n!}\left(\int\limits_{-\infty}^{+\infty}d ^{4}y\ \big[\psi\left(y\right) \gamma^{\beta}\star\overline\psi\left(y\right)\big]\ A_{\beta}\left(y\right)\right)^{n}. 
\end{eqnarray}
The contraction of the fermion fields is given by diagrams of Fig. [2]. It turns out that the contribution of the first diagram [2a] vanishes, whereas the next three diagrams lead to the field strength tensor:
\begin{eqnarray}\label{F338e}
F_{\beta\sigma}=\partial_{\beta}A_{\sigma}-\partial_{\sigma}A_{\beta}+ig\ [A_{\beta}, A_{\sigma}]_{\star}.
\end{eqnarray}
In the following we will show that the derivative terms in the above definition come from the diagram in Fig. [2b], whereas other diagrams in Figs. [2c] and [2d] yield the Moyal bracket of the background gauge fields. 
\par
Let us begin the analysis with the zeroth order of perturbation theory. The integral appearing from the contraction of the fermion-antifermion fields in the case $n=0$ is highly divergent in the limit $\epsilon \to 0$, but gives zero when traced with $\gamma^{\mu}\gamma^{5}$. This is the same result as in the commutative four-dimensional massless QED.  
\par
The first perturbative correction of ${\cal{T}}\left(x,\epsilon\right)$ is:  
\begin{eqnarray}\label{F339e}
{\cal{T}}_{1}\left(x,\epsilon\right)&=&ig\int\limits_{-\infty}^{+\infty} \frac{d^{4}p}{\left(2\pi\right)^{4}}
\frac{d^{4}q}{\left(2\pi\right)^{4}}
\frac{d^{4}\ell}{\left(2\pi\right)^{4}} 
\prod_{i=1,2}\frac{d^{4}\ell_{i}}{\left(2\pi\right)^{4}}
\bigg<\tilde{\psi}\left(p\right)\gamma^{\mu}\gamma^{5}\tilde{\overline{\psi}}\left(q\right)\tilde{{\psi}}\left(\ell_{1}\right)\gamma^{\beta}\tilde{\overline{\psi}}\left(\ell_{2}\right) \bigg>\nonumber\\
&&\ \ \times  \ \int\limits_{-\infty}^{+\infty}d^{4}y_{1} A_{\beta}\left(\ell\right)e^{-i\ell y_{1}}\   \left(e^{-i\ell_{1}y_{1}}\star e^{i\ell_{2}y_{1}}\right)\ \left(e^{-ipx}\star e^{iqx}\right)\ e^{i\left(q+p\right)\frac{\epsilon}{2}}.
\end{eqnarray}
Note that the product between the exponentials with $x$ and the exponentials with $y_{1}$ is the ordinary product of functions. The definition of the $\star$-product in the momentum space leads to:
\begin{eqnarray}\label{F340e}
{\cal{T}}_{1}\left(x,\epsilon\right)&=& ig\int\limits_{-\infty}^{+\infty} \frac{d^{4}p}{\left(2\pi\right)^{4}}
\frac{d^{4}q}{\left(2\pi\right)^{4}}
\frac{d^{4}\ell}{\left(2\pi\right)^{4}} 
\prod_{i=1,2}\frac{d^{4}\ell_{i}}{\left(2\pi\right)^{4}}
\bigg<\tilde{\psi}\left(p\right)\gamma^{\mu}\gamma^{5}\tilde{\overline{\psi}}\left(q\right)\tilde{{\psi}}\left(\ell_{1}\right)\gamma^{\beta}\tilde{\overline{\psi}}\left(\ell_{2}\right) \bigg>\nonumber\\
&&\ \ \ \ \ \times  \left(2\pi\right)^{4}\ \delta^{4}\left(\ell_{2}-\ell_{1}-\ell\right)A_{\beta}\left(\ell\right)\ e^{i\left(q-p\right)x}\  e^{i\left(q+p\right)\frac{\epsilon}{2}} e^{\frac{i\theta_{\eta\kappa}}{2}\left(p^{\eta}q^{\kappa}+\ell_{1}^{\eta}\ell_{2}^{\kappa}\right)},
\end{eqnarray}
where the integration over $y_{1}$ is performed to yield the $\delta$-function. Contraction of the fermions removes  
the exponential function containing the noncommutativity parameter because of the antisymmetry of $\theta$. After a change of integration variables we get:
\begin{eqnarray}\label{F341e}
{\cal{T}}_{1}\left(x,\epsilon\right)=-ig\int\frac{d^{4}s}{\left(2\pi\right)^{4}}\frac{d^{4}\ell}{\left(2\pi\right)^{4}}\ \mbox{Tr}\bigg[\frac{i\SS}{s^{2}}\ \gamma^{\mu}\gamma^{5}\ \frac{i\left(\SS-\LS\right)}{\left(s-\ell\right)^{2}}\ \gamma^{\beta}\bigg]\ A_{\beta}\left(\ell\right)\ e^{-i\ell x}\ e^{is\epsilon}.
\end{eqnarray} 
To evaluate the limit $\epsilon\to 0$, the integrand can be expanded for large external gauge field momentum $\ell$. We therefore obtain:
\begin{eqnarray}\label{F342e}
{\cal{T}}_{1}\left(x,\epsilon\right)&=&-ig\mbox{Tr}\left(\gamma^{\alpha}\gamma^{\mu}\gamma^{5}\gamma^{\sigma}\gamma^{\beta}\right)\int\limits_{-\infty}^{+\infty}\frac{d^{4}\ell}{\left(2\pi\right)^{4}}\ \ell_{\sigma}A_{\beta}\left(x\right)\ e^{-i\ell x}\int\limits_{-\infty}^{+\infty}\frac{d^{4}s}{\left(2\pi\right)^{4}}\ \frac{s_{\alpha}e^{is\epsilon}}{s^{4}},\nonumber\\
&=&-\frac{ig\ \varepsilon^{\mu\alpha\beta\sigma}}{4\pi^{2}}\left(\partial_{\sigma} A_{\beta}\left(x\right)\right)\frac{\partial}{\partial \epsilon_{\alpha}}\ \mbox{log}\frac{1}{\epsilon^{2}}\nonumber\\
&=&-\frac{ig\ \varepsilon^{\mu\alpha\beta\sigma}}{4\pi^{2}}\left(\partial_{\beta}A_{\sigma}\left(x\right)-\partial_{\sigma}A_{\beta}\left(x\right)\right)\ \frac{\epsilon_{\alpha}}{\epsilon^{2}}.
\end{eqnarray}
Here, we have used Tr $\left(\gamma^{5}\gamma^{\mu}\gamma^{\alpha}\gamma^{\beta}\gamma^{\sigma}\right)=-4i\varepsilon^{\mu\alpha\beta\sigma}$ in four dimensions. Again a finite result can be obtained by inserting this expression in the Eq. (\ref{F312}), now in four dimensions. In the first order of perturbative expansion the contribution to the divergence of the axial vector current is therefore given by:
\begin{eqnarray}\label{F343e}
\lefteqn{\hspace{-2cm}\bigg[\int \limits_{-\infty}^{+\infty}d^{4}x\ \bigg<\partial_{\mu}J^{\mu\left(5\right)}\left(x\right)\bigg>\bigg]_{\mbox {\small for $n=1$}}=
-ig\int\limits_{-\infty}^{+\infty}d^{4}x\ F_{\mu\nu}\left(x\right)\lim\limits_{\epsilon\to 0}\epsilon^{\nu}\ {\cal{T}}_{1}\left(x,\epsilon\right)}\nonumber\\
&=&-\frac{g^{2}}{4\pi^{2}}
\left(\lim\limits_{\epsilon\to 0}\ \frac{\epsilon_{\alpha}\epsilon^{\nu}}{\epsilon^{2}}\right) 
 \varepsilon^{\mu\alpha\beta\sigma} \int\limits_{-\infty}^{+\infty}d^{4}x\ \big[\partial_{\beta}A_{\sigma}\left(x\right)-\partial_{\sigma}A_{\beta}\left(x\right)\big]\ F_{\mu\nu}\left(x\right)\nonumber\\
&=&
-\frac{g^{2}}{16\pi^{2}} \varepsilon^{\mu\nu\beta\sigma} \int\limits_{-\infty}^{+\infty}d^{4}x\ \big[\partial_{\beta}A_{\sigma}\left(x\right)-\partial_{\sigma}A_{\beta}\left(x\right)\big]\ F_{\mu\nu}\left(x\right),
\end{eqnarray}
where we have made use of the Eq. (\ref{F315}) for $D=4$. In commutative QED the expression in the square brackets on the last line of the above result is the Abelian field strength tensor and the above result is  the familiar result for the axial anomaly obtained in the point-splitting method. However, it is  only after considering the contribution of the next order of perturbation theory, that the non-linear term in the definition of the field strength tensor in NC-QED can be recovered. 
\par
The contribution of the second order perturbation theory to ${\cal{T}}\left(x,\epsilon\right)$  can be obtained in a similar way.  The contraction of fermions yields two connected diagrams [Fig. [2c] and [2d]],
\begin{eqnarray}\label{F344e}
\lefteqn{{\cal{T}}_{2}\left(x,\epsilon\right)=-\frac{ig^{2}}{2!}\int\limits_{-\infty}^{+\infty}\frac{d^{4}k}{\left(2\pi\right)^{4}}\ \frac{d^{4}\ell}{\left(2\pi\right)^{4}}\ \frac{d^{4}s}{\left(2\pi\right)^{4}}
\ A_{\sigma}\left(k\right)\ A_{\beta}\left(\ell\right)\ 
 e^{-i\left(\ell+k\right)x}\ e^{is\epsilon}
}\nonumber\\
&&\times \Bigg\{ \mbox{Tr}\bigg[\frac{\SS+\KS}{\left(s+k\right)^{2}}\gamma_{\mu}\gamma_{5}\frac{\SS-\LS}{\left(s-\ell\right)^{2}}\gamma_{\beta}\frac{\SS}{s^{2}}\gamma_{\sigma}\bigg]
e^{-\frac{i\theta_{\eta\kappa}}{2}k^{\eta}\ell^{\kappa}}
+
[\left(\ell,\beta\right)\leftrightarrow\left(k,\sigma\right)] \Bigg\},
\end{eqnarray}
where the first expression on the last line belongs to the diagram in Fig. [2c] and $[\left(\ell,\beta\right)\leftrightarrow\left(k,\sigma\right)]$ denotes the contribution of diagram [2d]. Note that the effective phase $e^{-\frac{i\theta_{\eta\kappa}}{2}k^{\eta}\ell^{\kappa}}$ can be obtained after performing all the $\star$-products appearing in this order and considering all the $\delta$-functions of the momenta, which arise after contractions of fermion-antifermion fields.  
\par
Next we have to evaluate the trace 
$
\mbox{Tr}\big[\left(\SS+\KS\right)\gamma_{\mu}\gamma_{5}\left(\SS-\LS\right)\gamma_{\beta}\SS\gamma_{\sigma}\big]$ in the contribution of diagram [2c] and 
$
\mbox{Tr}\big[\left(\SS+\LS\right)\gamma_{\mu}\gamma_{5}\left(\SS-\KS\right)\gamma_{\sigma}\SS\gamma_{\beta}\big]
$ in the contribution of diagram [2d]. 
It turns out that the relevant contribution for the limit $\epsilon \to 0$ comes only from $\mbox{Tr}\left(\SS\gamma_{\mu}\gamma_{5}\SS\gamma_{\beta}\SS\gamma_{\sigma}\right)$ in both traces. Other terms lead to non-covariant expressions like $\big[\partial_{\tau}A_{\beta},A_{\sigma}\big]_{\star}\mbox{log}\epsilon^{2}$, which vanish after contracting with $-ig\epsilon^{\nu}F_{\mu\nu}\left(x\right)$ and taking the limit $\epsilon \to 0$.   
\par
For large $k$ and $\ell$ the relevant contribution to the VEV ${\cal{T}}\left(x,\epsilon\right)$, in second order of perturbation theory in the limit $\epsilon\to 0$ is given by:
\begin{eqnarray}\label{F345e}
{\cal{T}}_{2}\left(x,\epsilon\right)&=&
-\frac{ig^{2}}{2!}\bigg[\int\limits_{-\infty}^{+\infty}\frac{d^{4}k}{\left(2\pi\right)^{4}}\ A_{\sigma}\left(k\right)\ e^{-ik x}\star\int\limits_{-\infty}^{\infty}\frac{d^{4}\ell}{\left(2\pi\right)^{4}}\ 
e^{-i\ell x}A_{\beta}\left(\ell\right)\nonumber\\
&&\hspace{1.5cm}\times \int \frac{d^{4}s}{\left(2\pi\right)^{4}}e^{is\epsilon}\ \frac{\mbox{Tr}\left(\SS\gamma_{\mu}\gamma_{5}\SS\gamma_{\beta}\SS\gamma_{\sigma}\right) }{s^{6}} \Bigg]+[\left(\ell,\beta\right)\leftrightarrow\left(k,\sigma\right)]\nonumber\\
&=&
-\frac{ig^{2}}{2!}\ 
\Bigg[A_{\sigma}\left(x\right)\star 
A_{\beta}\left(x\right)
\int \frac{d^{4}s}{\left(2\pi\right)^{4}}e^{is\epsilon}\ \frac{\mbox{Tr}\left(\SS\gamma_{\mu}\gamma_{5}\SS\gamma_{\beta}\SS\gamma_{\sigma}\right) }{s^{6}} \Bigg]+[\beta\leftrightarrow\sigma],
\end{eqnarray}
where the definition of the $\star$-product is used on the first line to obtain $A_{\sigma}\star A_{\beta}$.
To evaluate the traces we can make use of the relation:
\begin{eqnarray}\label{F346e}
\mbox{Tr}\left(\gamma_{5}\gamma_{\rho}\gamma_{\beta}\gamma_{\kappa}\gamma_{\sigma}\gamma_{\tau}\gamma_{\mu}\right)&=&+4i\varepsilon_{\sigma\tau\mu\alpha}\left(\delta^{\alpha}_{\mu}g_{\beta\kappa}-\delta^{\alpha}_{\beta}g_{\rho\kappa}+\delta^{\alpha}_{\kappa}g_{\rho\beta}\right)\nonumber\\
&&-4i\varepsilon_{\rho\beta\kappa\alpha}\left(\delta^{\alpha}_{\sigma}g_{\tau\mu}-\delta^{\alpha}_{\tau}g_{\sigma\mu}+\delta^{\alpha}_{\mu}g_{\sigma\tau}\right),
\end{eqnarray}
to obtain:
\begin{eqnarray*}
\mbox{Tr}\left(\SS\gamma_{\mu}\gamma_{5}\SS\gamma_{\beta}\SS\gamma_{\sigma}\right)=-4i\varepsilon_{\sigma\tau\mu\beta}\ s_{\tau}\ s^{2}.
\end{eqnarray*}
The integration over $s$ in the Eq. (\ref{F343e}) can be performed in the same way as in Eq. (\ref{F342e}), giving: 
\begin{eqnarray}\label{F347e}
{\cal{T}}_{2}\left(x,\epsilon\right)=+\frac{g^{2}}{4\pi^{2}}\varepsilon^{\mu\tau\beta\sigma}\bigg[A_{\beta}\left(x\right),A_{\sigma}\left(x\right)\bigg]_{\star}\ \frac{\epsilon_{\tau}}{\epsilon^{2}}.
\end{eqnarray}
The contribution from the second order of perturbative expansion to the divergence of the axial current is therefore given by: 
\begin{eqnarray}\label{F348e}
\lefteqn{\hspace{-2cm}\bigg[\int \limits_{-\infty}^{+\infty}d^{4}x\ \bigg<\partial_{\mu}J^{\mu\left(5\right)}\left(x\right)\bigg>\bigg]_{\mbox {\small for $n=2$}}=
-ig\int\limits_{-\infty}^{+\infty}d^{4}x\ F_{\mu\nu}\left(x\right)\lim\limits_{\epsilon\to 0}\epsilon^{\nu}\ {\cal{T}}_{2}\left(x,\epsilon\right)}\nonumber\\
&=&
-\frac{ig^{3}}{4\pi^{2}} \left(\lim\limits_{\epsilon\to 0}\ \frac{\epsilon_{\tau}\epsilon^{\nu}}{\epsilon^{2}}\right)  \varepsilon^{\mu\tau\beta\sigma}\int\limits_{-\infty}^{+\infty}d^{4}x \bigg[A_{\beta}\left(x\right),A_{\sigma}\left(x\right)\bigg]_{\star}\ F_{\mu\nu}\left(x\right),\nonumber\\
&=&
-\frac{ig^{3}}{16\pi^{2}}\varepsilon^{\mu\nu\beta\sigma}\int\limits_{-\infty}^{+\infty}d^{4}x\ \bigg[A_{\beta}\left(x\right),A_{\sigma}\left(x\right)\bigg]_{\star}\ F_{\mu\nu}\left(x\right),
\end{eqnarray}
where we have made use of the Eqs. (\ref{F312}) now in four dimensions, and (\ref{F315}). Adding the result of the first order of expansion, Eq. (\ref{F343e}), to this result gives the axial anomaly in four-dimensional massless NC-QED:
\begin{eqnarray}\label{F349e}
\int \limits_{-\infty}^{+\infty}d^{4}x\ \bigg<\partial_{\mu}J^{\mu\left(5\right)}\left(x\right)\bigg>&=& -\frac{g^{2}}{16\pi^{2}}\varepsilon^{\mu\nu\beta\sigma}\int\limits_{-\infty}^{+\infty}d^{4}x\ F_{\mu\nu}\left(x\right)\star F_{\beta\sigma}\left(x\right).
\end{eqnarray}
Note that the integral over space-time coordinates plays the role of trace over group the indices in the commutative non-Abelian gauge theories, and guarantees the $\star$-gauge invariance of the result. Subsequently we will find the expression for the anomaly density in terms of Chern-Pointryagin density, {\it al beit} a non gauge invariant equation. 
\par\vskip0.3cm\par\noindent{\bf The Fujikawa Method}\label{S42}
\par\vskip0.3cm\par\noindent
In this method the axial anomaly is interpreted as a result of the non-invariance of the fermionic measure of integration under infinitesimal local axial gauge transformations. 
\par
We begin the evaluation of the axial anomaly by studying the transformation properties  of the partition function of NC-QED,
\begin{eqnarray}\label{F317}
{\cal{Z}}=\int D\psi\ D\overline{\psi}\ e^{-iS_{F}[\psi,\overline{\psi}]},
\end{eqnarray}
under local chiral gauge transformations:
\par\vskip0.5cm
\hspace{2cm}$\psi\left(x\right)\to\psi'\left(x\right)=U^{\left(5\right)}\left(x\right)\star\psi\left(x\right),\hspace{1cm}\overline{\psi}\left(x\right)\to \overline{\psi}'\left(x\right)=\overline{\psi}\left(x\right)\star U^{\left(5\right)}\left(x\right)$,
\par with
\begin{eqnarray}\label{F318}
U^{\left(5\right)}\left(x\right)=\left(e^{i\gamma_{5}\alpha\left(x\right)}\right)_{\star}.&
\end{eqnarray}
In the chiral limit and for vanishing background gauge field the fermionic action $S_{F}[\psi,\overline{\psi}]$, Eq. (\ref{F214}),
\begin{eqnarray}\label{F319}
S_{F}[\psi,\overline{\psi}]\equiv i\int d^{4}x\  \overline{\psi}\left(x\right)\gamma^{\mu}\star\partial_{\mu}\psi\left(x\right),
\end{eqnarray}
transforms as,
\begin{eqnarray}\label{F320}
S_{F}\to S_{F}'=S_{F}-\int d^{4}x \ \overline{\psi}\left(x\right)\star \partial_{\mu}\alpha\left(x\right) \ \gamma^{\mu}\gamma^{5}\star \psi\left(x\right).
\end{eqnarray} 
From the definition of the $\star$-product we get:
\begin{eqnarray*}
\int d^{4}x \ \overline{\psi}\left(x\right)\star \partial_{\mu}\alpha\left(x\right) \ \gamma^{\mu}\gamma^{5}\star \psi\left(x\right)=i \int d^{4}x\  \alpha\left(x\right)\star\partial_{\mu}J^{\mu \left(5\right)}\left(x\right),
\end{eqnarray*}
where the axial current $ J^{\mu\left(5\right)}$ is defined in the Eq. (\ref{F33}). The transformed fermionic action can  therefore be written as:
\begin{eqnarray}\label{F322}
S_{F}'[\psi',\overline{\psi}']=S_{F}[\psi,\overline{\psi}]-i\int d^{4}x\ \alpha\left(x\right)\star\partial_{\mu}J^{\mu\left(5\right)}\left(x\right).
\end{eqnarray}
Replacing this equation in the partition function of the fermion field yields,
\begin{eqnarray}\label{F323}
{\cal{Z}}'&=&\int D\psi' \ D\overline{\psi}'
e^{-iS_{F}'[\psi',\overline{\psi}']}\nonumber\\
&=&\int D\psi' \ D\overline{\psi}'\ e^{-iS_{F}[\psi,\overline{\psi}]}\ \mbox{exp}\bigg\{-\int d^{4}x
\ \alpha\left(x\right)\star\partial_{\mu}J^{\mu\left(5\right)}\left(x\right)
\bigg\}.
\end{eqnarray}
Under a general local matrix transformation $\psi\left(x\right)\to U\left(x\right)\star\psi\left(x\right)$, with $U\left(x\right)$ a unitary matrix, the integration measure of the fermionic fields will transform with the inverse of the determinants of the transform matrix\cite{weinberg}:
\begin{eqnarray*}
D\overline\psi\ D\psi\to \left(\mbox{Det}\  {\cal{U}}\ \mbox{Det}\ {\overline{\cal{U}}} \right)^{-1}D\overline{\psi}\ D\psi,
\end{eqnarray*} 
where
\begin{eqnarray}\label{F324}
{\cal{U}}_{x,y}\equiv U\left(x\right)\star\delta^{4}\left(x-y\right),\hspace{1.5cm}{\overline{\cal{U}}}_{x,y}\equiv \gamma_{0}U^{\dagger}\left(x\right)\gamma_{0}\star\delta^{4}\left(x-y\right).
\end{eqnarray}
Notice that in the noncommutative field theory the $\delta^{4}\left(x-y\right)$-function is definied as in the commutativ case, i.e. by $\int g\left(x\right)\star\delta^{4}\left(x-y\right) d^{4}x =g\left(y\right)$ for a test function $g\left(x\right)$. 
For a local chiral gauge transformation, replacing $U\left(x\right)$ in the (\ref{F324}) by the unitary matrix $U^{\left(5\right)}\left(x\right)$ from the Eq. (\ref{F318}), we obtain:
\begin{eqnarray}\label{F325}
{\cal{U}}_{x,y}=\overline{\cal{U}}_{x,y}.
\end{eqnarray}
The transformation rule for the fermionic integration measure reads then:
\begin{eqnarray}\label{F326}
D\psi\ D\overline{\psi}\to D\psi'\ D\overline{\psi}'=\left(\mbox{Det}\ {\cal{U}}_{x,y}\right)^{-2}D\psi\ D\overline{\psi}.
\end{eqnarray}
From the Eq. (\ref{F318}) the fermionic fields transform as:
\begin{eqnarray}\label{F327}
\overline{\psi}\left(x\right)\approx\overline{\psi}\left(x\right)+i\gamma_{5}  \overline{\psi}\left(x\right)\star \alpha\left(x\right),\hspace{1.5cm}
\psi\left(x\right)\approx\psi\left(x\right)+i\gamma_{5}\ \alpha\left(x\right)\star\psi\left(x\right),
\end{eqnarray} 
and the fermionic integration measure tranforms as:
\begin{eqnarray}\label{F328}
D\psi'\ D\overline{\psi}'&\approx& \left(\mbox{Det}\ {\cal{U}}_{x,y}\right)^{-2}\ D\psi\ D\overline{\psi}\nonumber\\
&\approx& \mbox{exp}\bigg[{i}\int d^{4}x\ \alpha\left(x\right)\star{\cal{
A}}\left(x\right)\bigg]\ D\psi\ D\overline{\psi}.
\end{eqnarray}  
with 
\begin{eqnarray}\label{F329}
{\cal{A}}\left(x\right)\equiv -2\ \mbox{Tr}\left(\gamma_{5}\right)\delta^{4}\left(x-x\right).
\end{eqnarray}
In this expression the $\delta$-function is infinite, whereas the trace over $\gamma_{5}$ vanishes. To regularize it in a $\star$-gauge {\it covariant}  manner a differentiable operator $f\left(-\frac{D_{\mu}\gamma^{\mu}\star D_{\nu}\gamma^{\nu}}{M^{2}}\right)$, 
with $D_{\mu}$ introduced in the Eq. (\ref{F215}), is inserted. Here, $M$ is a large mass, which should be taken eventually to infinity.  The operator $D_{\mu}\star D_{\nu}$ transforms in the usual covariant manner: $U\left(x\right)\star D_{\mu}\star D_{\nu}\star U^{-1}\left(x\right)$. This operator should act first on $\delta^{4}\left(x-y\right)$ before  taking the limit $x\to y$. The regularized anomaly function ${\cal{A}}\left(x\right)$ can therefore be given by:
\begin{eqnarray}\label{F330}
{\cal{A}}\left(x\right)=-2\ \mbox{Tr}\bigg\{\gamma_{5}f\left(\frac{-D_{\mu}\gamma^{\mu}\star D_{\nu}\gamma^{\nu}}{M^{2}}\right)\bigg\}\star \delta^{4}\left(x-y\right).
\end{eqnarray}
The Fourier component of the above function $\tilde{f}\left(k^{2}\right)$ should satisfy the following conditions:
\begin{eqnarray}\label{F331}
\tilde{f}\left(k^{2}=0\right)=1,&\hspace{2cm}&\tilde{f}\left(k^{2}\to \infty\right)=0,\nonumber\\
k^{2}\tilde{f}'\left(k^{2}\right)=0&\mbox{for}&k^{2}=0\hspace{0.5cm}\mbox{and for } k^{2}\to \infty.
\end{eqnarray}
Using 
\begin{eqnarray}\label{F332}
-D_{\mu}\gamma^{\mu}\star D_{\nu}\gamma^{\nu}=-D_{\mu}\star D^{\mu}-\frac{g}{2}\ \sigma_{\mu\nu}F^{\mu\nu}\left(x\right),
\end{eqnarray}
where $\sigma_{\mu\nu}\equiv \frac{i}{2}[\gamma_{\mu},\gamma_{\nu}]$ and 
$igF_{\mu\nu}\equiv [D_{\mu},D_{\nu}]_{\star}$, and  where the antisymmetry of $\sigma_{\mu\nu}$ leads to  $\sigma_{\mu\nu}D^{\mu}\star D^{\nu}=\frac{1}{2}\sigma_{\mu\nu}[D^{\mu},D^{\nu}]_{\star}$, we get:
\begin{eqnarray}\label{F333}
{\cal{A}}\left(x\right)=-2\lim\limits_{y\to x}\int\ \frac{d^{D}k}{\left(2\pi\right)^{D}}\mbox{Tr}\bigg[\gamma_{5}f\left(\frac{-D_{\mu}\star D^{\mu}-\frac{g}{2}\ \sigma_{\mu\nu}F^{\mu\nu}\left(x\right)}{M^{2}}\right)\bigg] \star e^{ik\left(x-y\right)}.
\end{eqnarray}
Next a Taylor expansion on the regulator function $f$ for large $M$ can be performed to yield:
\begin{eqnarray}\label{F334}
\lefteqn{\hspace{-0cm}
f\left(-\frac{D_{\mu}\star D^{\mu}}{M^{2}}+\frac{g}{2M^{2}}\sigma_{\mu\nu}F^{\mu\nu}\left(x\right)\right)=
}\nonumber\\
&&+\frac{g^{2}}{2!\times 4M^{4}}\left(\sigma_{\mu\nu}F^{\mu\nu}\left(x\right)\right)\star\left(\sigma_{\xi\rho}F^{\xi\rho}\left(x\right)\right)f''\left(\frac{-D_{\mu}\star D^{\mu}}{M^{2}}\right)+{\cal{O}}\left(\frac{1}{M^{6}}\right),
\end{eqnarray}
The first two terms in the Taylor expansion vanish due to their traces with a $\gamma_{5} $ matrix in ${\cal{A}}\left(x\right)$ [Eq. (\ref{F333})]. The only term which contributes is quadratic in $F_{\mu\nu}$.  It turns out that only the term with power $1/M^4$ leads to a mass independent expression which is just the expected axial anomaly.  Making use of the expression:
\begin{eqnarray*}
\mbox{Tr}\left(\gamma_{5}\sigma_{\mu\nu}\ \sigma_{\xi\rho}\right)=-4i\varepsilon_{\mu\nu\xi\rho},
\end{eqnarray*}
the anomaly function ${\cal{A}}\left(x\right)$, Eq. (\ref{F333}), is then given by:
\begin{eqnarray}\label{F335}
{\cal{A}}\left(x\right)=\frac{ig^{2}}{M^{4}}\varepsilon_{\mu\nu\xi\rho}F^{\mu\nu}\left(x\right)\star F^{\xi\rho}\left(x\right)\lim\limits_{y\to x}\int\frac{d^{D}k}{\left(2\pi\right)^{D}}\bigg[\tilde{f}''\left(\frac{-D_{\mu}\star D^{\mu}}{M^{2}}\right)\bigg]\star e^{ik\left(x-y\right)}.
\end{eqnarray} 
For vanishing background gauge field the above integration can be performed. Using
\begin{eqnarray}\label{F336}
\lim\limits_{x\to y} \tilde{f}''\left(\frac{-\partial_{\mu}^{2}}{M^{2}}\right)\star e^{ik\left(x-y\right)}={\tilde{f}}''\left(\frac{k^{2}}{M^{2}}\right),
\end{eqnarray}
we obtain:
\begin{eqnarray*}
\int\frac{d^{D}k}{\left(2\pi\right)^{D}}\ \tilde{f}''\left(\frac{k^{2}}{M^{2}}\right)=\frac{M^{4}}{16\pi^{2}}.
\end{eqnarray*}
The anomaly function thus reads:
\begin{eqnarray}\label{F337}
{\cal{A}}\left(x\right)=\frac{ig^{2}}{16\pi^{2}}\varepsilon_{\mu\nu\xi\rho}\ F^{\mu\nu}\left(x\right)\star F^{\xi\rho}\left(x\right). 
\end{eqnarray}
Under an infinitesimal local chiral gauge transformation the integration measure for the fermionic fields behaves as [see Eq. (\ref{F328})]: 
\begin{eqnarray*}
D\psi'\ D\overline{\psi}'\approx \mbox{exp}\bigg[i\int d^{4}x\ \alpha\left(x\right)\star {\cal{A}}\left(x\right)\bigg]D\psi \ D\overline{\psi},
\end{eqnarray*}
with ${\cal{A}}\left(x\right)$ given in the Eq. (\ref{F337}). By making use of this transformation law the transformed partition function ${\cal{Z}}'$ from the Eq. (\ref{F323}) can be completed to yield:
\begin{eqnarray}\label{F338}
{\cal{Z}}'&=&\int D\psi' \ D\overline{\psi}'
e^{-iS_{F}'[\psi',\overline{\psi}']}\nonumber\\
&=&\int D\psi \ D\overline{\psi}\ e^{-iS_{F}[\psi,\overline{\psi}]}\ \mbox{exp}\bigg\{\int d^{4}x
\bigg[\alpha\left(x\right)\star\left(i{\cal{A}}\left(x\right)-\partial_{\mu}J^{\mu\left(5\right)}\left(x\right)\right)\bigg]
\bigg\}.
\end{eqnarray}
The $\star$-product in the integration over $x$ can be neglected. To derive the axial anomaly, we make use of the invariance of the partition function under any infinitesimal change of variables. The axial anomaly reads therefore:
\begin{eqnarray}\label{F339}
<\partial_{\mu}J^{\mu\left(5\right)}\left(x\right)>=-\frac{g^{2}}{16\pi^{2}}\varepsilon_{\mu\nu\xi\rho}F^{\mu\nu}\left(x\right)\star F^{\xi\rho}\left(x\right),
\end{eqnarray} 
which is only covariant under $\star$-gauge transformation though (\ref{F338}) is in fact gauge invariant. This is due to the covariant regularization performed in the Eq. (\ref{F330}). Note that $\star$-gauge invariance is guaranteed only by integrating the above result over $x$.   
\par\vskip0.3cm\par\noindent{\bf Triangle Anomaly}\label{S43}\par\vskip0.3cm\par\noindent
We will now present the direct calculation of the axial anomaly\cite{adler} by calculating the triangle diagrams. As a consequence of noncommutativity of QED, phases  will appear in the Feynman integrals corresponding to triangle diagrams. These phases will turn out to be independent of the internal loop momentum, however. Hence no nonplanar triangle diagrams appear in this case, as it was also originally suggested in  Ref.\cite{hayakawa2}.
\par
Let us begin the calculation with the three-point function of two vector and one axial vector currents:
\begin{eqnarray}\label{F373e}
\Gamma^{\mu\lambda\nu}\left(x,y,z\right)=\bigg<\mbox{T}\ \left(J^{\mu\left(5\right)}\left(x\right)J^{\lambda}\left(y\right)J^{\nu}\left(z\right)\right)\bigg>.
\end{eqnarray}
The axial vector current $J^{\mu\left(5\right)}$ is defined in the Eq. (\ref{F33}) and the vector current $J^{\mu}$ is given by:
\begin{eqnarray}\label{F374e}
J^{\mu}\left(x\right)=ig\ \overline{\psi}\left(x\right)\gamma^{\mu}\star\psi\left(x\right).
\end{eqnarray}
The operation between the currents in Eq. (\ref{F373e}) is just the usual product of functions,  
\begin{eqnarray}\label{F375e}
\lefteqn{\hspace{-0.5cm}\Gamma^{\mu\lambda\nu}\left(x,y,z\right)=-ig^{2}
\int\limits_{-\infty}^{+\infty}\prod_{i=1,2,3}\frac{d^{4}p_{i}}{\left(2\pi\right)^{4}}  
\prod_{j=1,2,3}\frac{d^{4}p'_{j}}{\left(2\pi\right)^{4}}\ 
e^{i\left(p_{1}-p'_{1}\right)x}e^{i\left(p_{2}-p'_{2}\right)y}e^{i\left(p_{3}-p'_{3}\right)z}}\nonumber\\
&&\hspace{-0.5cm}\Bigg< \tilde{\overline{\psi}}\left(p_{1}\right)
\gamma^{\mu}\gamma^{5}
\tilde{\psi}\left(p'_{1}\right)
\tilde{\overline{\psi}}\left(p_{2}\right)
\gamma^{\lambda}
\tilde{\psi}\left(p'_{2}\right)
\tilde{\overline{\psi}}\left(p_{3}\right)
\gamma^{\nu}
\tilde{\psi}\left(p'_{3}\right)
\Bigg> e^{\frac{i\theta_{\eta\sigma}}{2}\sum_{j=1,2,3}\ p_{j}^{\eta}p_{j}^{'\sigma}}.
\end{eqnarray}
There are  two Feynman diagrams [Fig. [3]] contributing to this expression,  with the result:
\begin{eqnarray}\label{F376e}
\lefteqn{\hspace{-1cm}\Gamma^{\mu\lambda\nu}\left(x,y,z\right)=-ig^{2}\int\limits_{-\infty}^{+\infty}
\frac{d^{4}k_{2}}{\left(2\pi\right)^{4}}
\frac{d^{4}k_{3}}{\left(2\pi\right)^{4}}
\frac{d^{4}\ell}{\left(2\pi\right)^{4}}
e^{-i\left(k_{2}+k_{3}\right)x}\ e^{ik_{2}y}e^{ik_{3}z}
}\nonumber\\
&&\hspace{-1.7cm} \times \Bigg\{\bigg[\mbox{Tr}\left(\frac{1}{D\left(\ell-k_{3}\right)}\gamma^{\mu}\gamma^{5}\frac{1}{D\left(\ell+k_{2}\right)}\gamma^{\lambda}\frac{1}{D\left(\ell\right)}\gamma^{\nu}\right)e^{+\frac{i\theta_{\eta\sigma}}{2}k_{2}^{\eta}k_{3}^{\sigma}}\bigg]+ \bigg[\left(k_{2},\lambda\right)\leftrightarrow\left(k_{3},\nu\right)\bigg] \Bigg\},
\end{eqnarray}
where the first expression on the second line is the contribution of diagram [3a], whereas $[\left(k_{2},\lambda\right)\leftrightarrow\left(k_{3},\nu\right)]$ denotes the contribution of diagram [3b]. Here $D\left(\ell\right)$ is the denominator of the fermion propagator  and is defined by 
$D\left(\ell\right)\equiv\LS-m$. As we can see, the phases $e^{\pm \frac{i\theta_{\eta\sigma}}{2} k_{2}^{\eta}k_{3}^{\sigma}}$ corresponding to both triangle diagrams [3a] and [3b] are  independent of the loop momentum $\ell$. Hence no nonplanar diagrams appear in this zeroth order of perturbative expansion.  
\par\vskip0.5cm
\noindent{\it i) Vector Ward Identity}
\par\noindent
To verify the vector Ward identities, we consider $\partial_{\lambda}\Gamma^{\mu\lambda\nu}$, and use the identity 
\begin{nedalph}\label{F378ea}
\KS_{2}=D\left(\ell+k_{2}\right)-D\left(\ell\right),
\end{eqnarray}
in the contribution of diagram [3a] and 
\begin{eqnarray}\label{F378eb}
\KS_{2}=D\left(\ell\right)-D\left(\ell-k_{2}\right),
\end{nedalph}
in the contribution of diagram [3b], to obtain
\begin{eqnarray}\label{F379e}
\lefteqn{ \frac{\partial}{\partial y^{\lambda}}\Gamma^{\mu\lambda\nu}\left(x,y,z\right) =
g^{2} \int\limits_{-\infty}^{+\infty}
\frac{d^{4}k_{2}}{\left(2\pi\right)^{4}}
\frac{d^{4}k_{3}}{\left(2\pi\right)^{4}}
\frac{d^{D}\ell}{\left(2\pi\right)^{D}}
\ e^{-i\left(k_{2}+k_{3}\right)x}\ e^{ik_{2}y}\ e^{ik_{3}z}
}\nonumber\\
&&\hspace{-0.3cm}\times 
\Bigg\{\Bigg[\mbox{Tr}\left(
\frac{1}{D\left(\ell-k_{3}\right)}\gamma^{\mu}\gamma^{5}\frac{1}{D\left(\ell\right)}\gamma^{\nu}\right)-\mbox{Tr}\left(\frac{1}{D\left(\ell-k_{3}\right)}\gamma^{\mu}\gamma^{5}\frac{1}{D\left(\ell+k_{2}\right)}\gamma^{\nu}\right)\Bigg]\ e^{+\frac{i\theta_{\eta\sigma}}{2}k_{2}^{\eta}k_{3}^{\sigma}}
\nonumber\\
&&+  \Bigg[\mbox{Tr}\left(
\frac{1}{D\left(\ell-k_{2}\right)}\gamma^{\mu}\gamma^{5}\frac{1}{D\left(\ell+k_{3}\right)}\gamma^{\nu}\right)-\mbox{Tr}\left(\gamma^{\mu}\gamma^{5}\frac{1}{D\left(\ell+k_{3}\right)}\gamma^{\nu}\frac{1}{D\left(\ell\right)}\right)\Bigg]\ e^{-\frac{i\theta_{\eta\sigma}}{2}k_{2}^{\eta}k_{3}^{\sigma}}\Bigg\}.\nonumber\\
\end{eqnarray}
The integrands on the second and third lines are linearly divergent in the limit $D\to 4$. In the usual commutative QED, where no phases appear, these integrals could be shown to cancel after appropriate shifts of integration variables, which are justified by dimensional regularization. In NC-QED, however, no cancellation takes place because of different phases in diagrams [3a] and [3b]. After a shift $\ell\to \ell+k_{3}$ and $\ell\to \ell+k_{3}-k_{2}$ in the first and second integrands respectively, we arrive at:
\begin{eqnarray}\label{F380e}
\lefteqn{
\frac{\partial}{\partial y^{\lambda}}\Gamma^{\mu\lambda\nu}\left(x,y,z\right)=2ig^{2} \int\limits_{-\infty}^{+\infty}
\frac{d^{4}k_{2}}{\left(2\pi\right)^{4}}
\frac{d^{4}k_{3}}{\left(2\pi\right)^{4}}
\ e^{-i\left(k_{2}+k_{3}\right)x}\ e^{ik_{2}y}\ e^{ik_{3}z} \sin\left(\frac{\theta_{\eta\sigma}}{2}k_{2}^{\eta}k_{3}^{\sigma}\right)    
}\nonumber\\
&&\hspace{-.5cm}\times\int\limits_{-\infty}^{+\infty} \frac{d^{D}\ell}{\left(2\pi\right)^{D}}\Bigg[\mbox{Tr}\left(\frac{1}{D\left(\ell\right)}\gamma^{\mu}\gamma^{5}\frac{1}{D\left(\ell+k_{3}\right)}\gamma^{\nu}\right)-\mbox{Tr}\left(\frac{1}{D\left(\ell-k_{2}\right)}\gamma^{\mu}\gamma^{5}\frac{1}{D\left(\ell+k_{3}\right)}\gamma^{\nu}\right)\Bigg]. \nonumber\\
\end{eqnarray}
Integration over the loop momentum $\ell$ vanishes due to the symmetry properties of the trace. We therefore obtain:
\begin{eqnarray}\label{F381e}
\frac{\partial}{\partial y^{\lambda}}\Gamma^{\mu\lambda\nu}\left(x,y,z\right)=0.
\end{eqnarray}
The triangle graphs make a contribution to the current in the presence of a gauge field $A_{\nu}$ and an axial gauge field $A_{\mu}^{\left(5\right)}$:
\begin{eqnarray}\label{F382e}
\big<J^{\lambda}\left(y\right)\big>\equiv\frac{1}{2}\int d^{4}x\  d^{4}z\  \Gamma^{\mu\lambda\nu}\left(x,y,z\right)A_{\mu}^{\left(5\right)}\left(x\right)A_{\nu}\left(z\right).
\end{eqnarray}
By making use of the Eq. (\ref{F381e}) we therefore find the standard vector Ward identity in NC-QED:
\begin{eqnarray}\label{F383ea}
\big<\partial_{\lambda}J^{\lambda}\left(y\right)\big>=0.
\end{eqnarray}
\par\noindent{\it ii) Axial Vector Ward Identity}
\par\noindent
To evaluate the divergence of the axial vector current,  we consider:
\begin{eqnarray}\label{F384e}
\lefteqn{\hspace{-2cm}\frac{\partial}{\partial x^{\mu}}\Gamma^{\mu\lambda\nu}\left(x,y,z\right)=-g^{2}\int\limits_{-\infty}^{+\infty}
\frac{d^{4}k_{2}}{\left(2\pi\right)^{4}}
\frac{d^{4}k_{3}}{\left(2\pi\right)^{4}}
\frac{d^{D}\ell}{\left(2\pi\right)^{D}}
e^{-i\left(k_{2}+k_{3}\right)x}\ e^{ik_{2}y}e^{ik_{3}z}
}\nonumber\\
&&\hspace{-1.2cm} \times \Bigg\{\bigg[\mbox{Tr}\left(\frac{1}{D\left(\ell-k_{3}\right)}\left(\KS_{2}+\KS_{3}\right)\gamma^{5}\frac{1}{D\left(\ell+k_{2}\right)}\gamma^{\lambda}\frac{1}{D\left(\ell\right)}\gamma^{\nu}\right)e^{+\frac{i\theta_{\eta\sigma}}{2}k_{2}^{\eta}k_{3}^{\sigma}}\bigg]\nonumber\\
&&
+\bigg[\left(k_{2},\lambda\right)\leftrightarrow\left(k_{3},\nu\right)\bigg]\Bigg\}.
\end{eqnarray}
We follow the standard definition of $\gamma^{5}$-matrix  in $D$-di\-men\-sions (\cite{thooft}, \cite{peskin}):
\begin{eqnarray*}
\gamma^{5}=i\gamma^{0}\gamma^{1}\gamma^{2}\gamma^{3},
\end{eqnarray*}
where $\gamma^{5}$ anticommutes with $\gamma^{\mu}$ for $\mu=0,1,2,3$ but commutes with $\gamma^{\mu}$ for other values of $\mu$. The loop momentum will have components in all dimensions, whereas the external momenta $k_{2}$ and $k_{3}$ are still defined only in four dimensions. Using the separation $\ell=\ell_{||}+\ell_{\perp}$, where $\ell_{||}$ has nonzero components in dimensions $0,1,2,3$ and $\ell_{\perp}$ has nonzero components in the other $D-4$ dimensions, and  the identity:
\begin{nedalph}\label{F386ea}
\left(\KS_{2}+\KS_{3}\right)\gamma^{5}=-\gamma^{5}D\left(\ell+k_{2}\right)-D\left(\ell-k_{3}\right)\gamma^{5}-2m\gamma^{5}+2\gamma^{5}\LS_{\perp},
\end{eqnarray}
in the first expression on the second line of Eq. (\ref{F384e}), and  
\begin{eqnarray}\label{F386eb}
\left(\KS_{2}+\KS_{3}\right)\gamma^{5}=-\gamma^{5}D\left(\ell+k_{3}\right)-D\left(\ell-k_{2}\right)\gamma^{5}-2m\gamma^{5}+2\gamma^{5}\LS_{\perp},
\end{nedalph}
in the last expression on the second line of this equation, we obtain:
\begin{nedalph}\label{F387ea}
\lefteqn{\hspace{-1.5cm} \frac{\partial}{\partial x^{\mu}}\Gamma^{\mu\lambda\nu}\left(x,y,z\right)=
\int\limits_{-\infty}^{+\infty}
\frac{d^{4}k_{2}}{\left(2\pi\right)^{4}}
\frac{d^{4}k_{3}}{\left(2\pi\right)^{4}}\ e^{-i\left(k_{2}+k_{3}\right) x}
\ e^{ik_{2}y}\ e^{ik_{3}z}
}\nonumber\\
&&\times \bigg[2mg^{2}T^{\lambda\nu}\left(m,k_{2},k_{3};\theta\right)+A^{\lambda\nu}\left(k_{2},k_{3};\theta\right)+R^{\lambda\nu}\left(m,k_{2},k_{3};\theta\right)\bigg],
\end{eqnarray}
where
\begin{eqnarray}\label{F387eb}
T^{\lambda\nu}\left(m,k_{2},k_{3};\theta\right)&\equiv&
\int\limits_{-\infty}^{+\infty}\frac{d^{D}\ell}{\left(2\pi\right)^{D}}
\Bigg[\mbox{Tr} \left(
\frac{1}{D\left(\ell-k_{3}\right)}\gamma^{5}\frac{1}{D\left(\ell+k_{2}\right)}\gamma^{\lambda}\frac{1}{D\left(\ell\right)}\gamma^{\nu}\right)
e^{+\frac{i\theta_{\eta\sigma}}{2}k_{2}^{\eta}k_{3}^{\sigma}}
\Bigg]\nonumber\\
&&+\bigg[\left(k_{2},\lambda\right)\leftrightarrow \left(k_{3},\nu\right)\bigg],
\end{eqnarray}
and 
\begin{eqnarray}\label{F387ec}
\lefteqn{A^{\lambda\nu}\left(k_{2},k_{3};\theta\right)\equiv}\nonumber\\
&\equiv&-2g^{2} \int\limits_{-\infty}^{+\infty}\frac{d^{D}\ell}{\left(2\pi\right)^{D}}
\Bigg[\mbox{Tr}\left( \frac{1}{D\left(\ell-k_{3}\right)}\gamma^{5}\LS_{\perp}\frac{1}{D\left(\ell+k_{2}\right)}\gamma^{\lambda}\frac{1}{D\left(\ell\right)}\gamma^{\nu}\right)\ e^{\frac{+i\theta_{\eta\sigma}}{2}k_{2}^{\eta}k_{3}^{\sigma}}\Bigg]\nonumber\\
&&+\bigg[\left(k_{2},\lambda\right)\leftrightarrow \left(k_{3},\nu\right)\bigg],\end{eqnarray}
and 
\begin{eqnarray}\label{F387ed}
\lefteqn{\hspace{-1.5cm}R^{\lambda\nu}\left(m,k_{2},k_{3};\theta\right)\equiv g^{2}\int\limits_{-\infty}^{+\infty}\frac{d^{D}\ell}{\left(2\pi\right)^{D}}
\Bigg[\Bigg\{\mbox{Tr}\left(\frac{1}{D\left(\ell-k_{3}\right)}\gamma^{5}\gamma^{\lambda}\frac{1}{D\left(\ell\right)}\gamma^{\nu}\right)}\nonumber\\
&&+
\mbox{Tr}\left(\gamma^{5}\frac{1}{D\left(\ell+k_{2}\right)}\gamma^{\lambda}\frac{1}{D\left(\ell\right)}\gamma^{\nu}\right)\Bigg\}
e^{+\frac{i\theta_{\eta\sigma}}{2}k_{2}^{\eta}k_{3}^{\sigma}}
\Bigg]+\bigg[\left(k_{2},\lambda\right)\leftrightarrow \left(k_{3},\nu\right)\bigg].
\end{nedalph}
Using the standard Feynman parametrization we obtain the following results for the above functions. For $T^{\lambda\nu}$ we get:
\begin{eqnarray}\label{F388e}
\lefteqn{\hspace{-1.5cm}T^{\lambda\nu}\left(m,k_{2},k_{3};\theta\right)=\frac{m}{2\pi^{2}}\varepsilon^{\lambda\nu\alpha\beta}k_{2\alpha}k_{3\beta}\cos\left(\frac{\theta_{\eta\sigma}}{2}k_{2}^{\eta}k_{3}^{\sigma}\right) }  \nonumber\\
&&\times\int\limits_{0}^{1}dx\ \int\limits_{0}^{1-x}dy\ \frac{1}{m^{2}-k_{2}^{2}x\left(1-x\right)-k_{3}^{2}y\left(1-y\right)-2k_{2}k_{3}xy}.
\end{eqnarray}
Note that the contribution of this function to $\partial_{\mu}\Gamma^{\mu\lambda\nu}\left(x,y,z\right)$ vanishes after taking the chiral limit $m\to 0$ in the Eq. (\ref{F387ea}).
\par
For the anomalous function $A^{\lambda\nu}$ we have: 
\begin{eqnarray}\label{F389e}
A^{\lambda\nu}\left(k_{2},k_{3};\theta\right)= -\frac{g^{2}}{2\pi^{2}}\varepsilon^{\lambda\nu\alpha\beta}k_{2\alpha}k_{3\beta}\cos\left(\frac{\theta_{\eta\sigma}}{2}k_{2}^{\eta}k_{3}^{\sigma}\right),
\end{eqnarray}
which is independent of the fermion mass $m$ and survives the chiral limit $m\to 0$. This is the anomalous contribution to $\partial_{\mu}\Gamma^{\mu\lambda\nu}\left(x,y,z\right)$. For vanishing noncommutativity parameter $\theta_{\eta\sigma}$ this result is exactly the  U(1)-anomaly in commutative QED. 
\par
For the rest term $R^{\lambda\nu}\left(m,k_{2},k_{3};\theta\right)$, after performing a shift of integration variable $\ell$, we obtain:
\begin{eqnarray}\label{F390e}
\lefteqn{R^{\lambda\nu}\left(m,k_{2},k_{3};\theta\right)=-2ig^{2}\sin\left(\frac{\theta_{\eta\sigma}}{2}k_{2}^{\eta}k_{3}^{\sigma}\right) 
}\nonumber\\
&&\hspace{-0.5cm}\times \int\limits_{-\infty}^{\infty}\frac{d^{D}\ell}{\left(2\pi\right)^{D}}  \Bigg[\mbox{Tr}\left(\frac{1}{D\left(\ell\right)}\gamma^{\lambda}\gamma^{5}\frac{1}{D\left(\ell+k_{3}\right)}\gamma^{\nu}\right)-\mbox{Tr}\left(\frac{1}{D\left(\ell\right)}\gamma^{\lambda}\gamma^{5}\frac{1}{D\left(\ell-k_{2}\right)}\gamma^{\nu}\right)\Bigg].
\end{eqnarray}
Here, a  power counting shows that both  integrals are linearly divergent in the limit $D\to 4$. But they vanish due to symmetry properties of the traces over the $\gamma$-matrices. We therefore have:
\begin{eqnarray}\label{F391e}
R^{\lambda\nu}\left(m,k_{2},k_{3};\theta\right)=0.
\end{eqnarray}
Inserting Eqs. (\ref{F388e})-(\ref{F391e}) in Eq. (\ref{F387ea}) and  taking the chiral limit $m\to 0$ we get:
\begin{eqnarray}\label{F392e}
\lefteqn{\Bigg[\frac{\partial}{\partial x^{\mu}}\Gamma^{\mu\lambda\nu}\left(x,y,z\right)\Bigg]_{m\to 0}=}\nonumber\\
&=&
-\frac{g^{2}}{2\pi^{2}} \varepsilon^{\lambda\nu\alpha\beta} \int\limits_{-\infty}^{+\infty}
\frac{d^{4}k_{2}}{\left(2\pi\right)^{4}}
\frac{d^{4}k_{3}}{\left(2\pi\right)^{4}}\ e^{-i\left(k_{2}+k_{3}\right) x}
\ e^{ik_{2}y}\ e^{ik_{3}z}
k_{2\alpha}k_{3\beta}\cos\left(\frac{\theta_{\eta\sigma}}{2}k_{2}^{\eta}k_{3}^{\sigma}\right).
\end{eqnarray}
Thus the contribution of triangle diagrams to axial vector current in the presence of vector gauge fields,
\begin{eqnarray}\label{F393e}
\big<J^{\mu\left(5\right)}\left(x\right)\big>\equiv\frac{1}{2}\int d^{4}y \ d^{4}z\ \Gamma^{\mu\lambda\nu}\left(x,y,z\right)\ A_{\lambda}\left(y\right)A_{\nu}\left(z\right),
\end{eqnarray}
and Eq. (\ref{F392e}) give,
\begin{eqnarray}\label{F394e} \bigg<\partial_{\mu}J^{\mu\left(5\right)}\left(x\right)\bigg>=-\frac{g^{2}}{4\pi^{2}} \varepsilon^{\lambda\nu\alpha\beta} \partial_{\lambda}A_{\nu}\left(x\right)\star\partial_{\alpha}A_{\beta}\left(x\right),
\end{eqnarray}
for the divergence of the axial vector current. To obtain the result by the point-splitting method [Eq. (\ref{F339})]:
\begin{eqnarray}\label{F395e}
\bigg<\partial_{\mu}J^{\mu\left(5\right)}\left(x\right)\bigg>= -\frac{g^{2}}{16\pi^{2}} \ \varepsilon^{\lambda\nu\alpha\beta} F_{\lambda\nu}\left(x\right)\star F_{\alpha\beta}\left(x\right),
\end{eqnarray}  
we have to consider the  contributions of all the diagrams shown in Figure [4]. Diagrams [4b] and [4c] turn out to be planar, so that no new effects will occur.  Note also that only integrating over the variable $x$ insures the $\star$-gauge invariance of the above result. 
\par\vskip0.3cm\noindent{\bf The Algebra of Currents and the Ward Identities}\par\vskip0.3cm\noindent
In what follows we will look anew at the Ward identities in NC-QED to show the analogy between noncommutative U(1) gauge theory and commutative non-Abelian gauge theories. In the previous section we have shown that as a consequence of the noncommutativity, phases will appear in the Feynman integrals for the triangle diagrams. In the commutative non-Abelian gauge theory these phases are replaced by traces over group generators\cite{weinberg}. A separation of these traces in an antisymmetric and a symmetric part leads to a separation between a ''formal'' part, which is connected to the algebra of currents and involve the structure constants of the group, and a ''potentially anomalous'' parts in the divergences of $\Gamma^{\mu\lambda\nu}$. The potentially anomalous part in the vector Ward identity can be chosen to vanish and only the anomalous part of axial vector Ward identity yields the usual axial anomaly. 
\par
In what follows we will show, that in NC-QED the same separation between the formal and the anomalous part in vector- and axial vector Ward identities is possible. We will use dimensional regularization to calculate the linearly divergent integrals appearing in the calculation. We will show that the formal part of both vector and axial vector Ward identity vanish after this regularization. Assuming gauge invariance, the anomalous part of the vector Ward identity can be chosen to vanish, and the anomalous part of the axial vector Ward identity to contain the anomaly.
\par\vskip1cm
\noindent{\it i) Vector Ward Identity}
\par\noindent
The starting point here is the three-point function $\Gamma^{\mu\lambda\nu}\left(x,y,z\right)$ given in the Eq. (\ref{F373e}). Using the definition of the time-ordered product appearing on the r.h.s. of this equation and deriving the resulting expression with respect to $y^{\lambda}$, we obtain the separation:
\begin{nedalph}\label{F396e}
\partial_{\lambda}\Gamma^{\mu\lambda\nu}\left(x,y,z\right)= 
\big[\partial_{\lambda}\Gamma^{\mu\lambda\nu}\left(x,y,z\right)\big]_{\mbox{\small{formal}}} +\big[\partial_{\lambda}\Gamma^{\mu\lambda\nu}\left(x,y,z\right)\big]_{\mbox{\mbox{\small {anomal}}}}, 
\end{eqnarray}
where the formal part is given by:
\begin{eqnarray}\label{F396eb}
\big[\partial_{\lambda}\Gamma^{\mu\lambda\nu}\left(x,y,z\right)\big]_{\mbox{\small{formal}}}& \equiv&\delta\left(y^{0}-z^{0}\right)\bigg<\mbox{T}\left(J^{\mu\left(5\right)}\left(x\right)\bigg[J^{0}\left(y\right),J^{\nu}\left(z\right)\bigg]\right)\bigg>\nonumber\\
&&\hspace{-0.5cm}+\delta\left(y^{0}-x^{0}\right)\bigg<\mbox{T}\left(\bigg[J^{0}\left(y\right),J^{\mu\left(5\right)}\left(x\right)\bigg]J^{\nu}\left(z\right)\right)\bigg>,
\end{eqnarray}
and the potentially anomalous part by:
\begin{eqnarray}\label{F396ec}
\big[\partial_{\lambda}\Gamma^{\mu\lambda\nu}\left(x,y,z\right)\big]_{\mbox{\small{anomal}}}\equiv\bigg<\mbox{T}\left(J^{\mu\left(5\right)}\left(x\right)\left(\partial_{\lambda}J^{\lambda}\left(y\right)\right)J^{\nu}\left(z\right)\right)\bigg>. 
\end{nedalph}
Note that, the commutators above are the usual commutators of operators:
\begin{eqnarray*}
[f\left(x\right),g\left(y\right)]\equiv f\left(x\right)g\left(y\right)-g\left(y\right)f\left(x\right),
\end{eqnarray*}  
and  the $\star$-products appear only in the definition of the vector and the axial vector current. Using the definition of the currents one obtains the following algebra of the vector and axial vector currents,   
\begin{nedalph}
\big[J^{0}\left(x\right),J^{\lambda}\left(y\right)\big]_{x_{0}=y_{0}}&=&-\left(J^{\lambda}\left(x\right)-J^{\lambda}\left(y\right)\right)\star \delta^{3}\left(x-y\right),\label{F397ea}\\
\big[J^{0\left(5\right)}\left(x\right),J^{\lambda\left(5\right)}\left(y\right)\big]_{x_{0}=y_{0}}&=&-\left(J^{\lambda}\left(x\right)-J^{\lambda}\left(y\right)\right)\star \delta^{3}\left(x-y\right),\label{F397eb}\\
\big[J^{0}\left(x\right),J^{\lambda\left(5\right)}\left(y\right)\big]_{x_{0}=y_{0}}&=&-\left(J^{\lambda\left(5\right)}\left(x\right)-J^{\lambda\left(5\right)}\left(y\right)\right)\star \delta^{3}\left(x-y\right),\label{F397ec}\\
\big[J^{0\left(5\right)}\left(x\right),J^{\lambda}\left(y\right)\big]_{x_{0}=y_{0}}&=&-\left(J^{\lambda\left(5\right)}\left(x\right)-J^{\lambda\left(5\right)}\left(y\right)\right)\star \delta^{3}\left(x-y\right).\label{F397ed}
\end{nedalph}
It is possible to compare these relations with equal-time commutation relations of QCD. There the structure constants $f_{abc}$ appear on the right-hand side of the commutation relations. In NC-QED, however, these structure constants are replaced by $2i\sin\left(\frac{\theta_{\eta\sigma}}{2}p^{\eta}q^{\sigma}\right)$ where $p$ and $q$ are the momenta of the first and second current in the commutation relation, respectively. 
\par
Using  Eqs. (\ref{F397ea}) and (\ref{F397ec}), the formal part of $\partial_{\lambda}\Gamma^{\mu\lambda\nu}$ reads:
\begin{eqnarray}\label{F398e} 
\lefteqn{
\Bigg[\frac{\partial}{\partial y^{\lambda}}\Gamma^{\mu\lambda\nu}\left(x,y,z\right)\Bigg]_{\mbox{\small{formal}}}=2ig^{2} \int\limits_{-\infty}^{+\infty}
\frac{d^{4}k_{2}}{\left(2\pi\right)^{4}}
\frac{d^{4}k_{3}}{\left(2\pi\right)^{4}}
\ e^{-i\left(k_{2}+k_{3}\right)x}\ e^{ik_{2}y}\ e^{ik_{3}z} \sin\left(\frac{\theta_{\eta\sigma}}{2}k_{2}^{\eta}k_{3}^{\sigma}\right)    
}\nonumber\\
&&\hspace{-.5cm}\times\int\limits_{-\infty}^{+\infty} \frac{d^{D}\ell}{\left(2\pi\right)^{D}}\Bigg[
\mbox{Tr}\left(\frac{1}{D\left(\ell\right)}\gamma^{\mu}\gamma^{5}\frac{1}{D\left(\ell+k_{3}\right)}\gamma^{\nu}\right) -
\mbox{Tr}\left(\frac{1}{D\left(\ell\right)}\gamma^{\mu}\gamma^{5}\frac{1}{D\left(\ell+k_{2}+k_{3}\right)}\gamma^{\nu}\right)\Bigg]
 ,\nonumber\\
\end{eqnarray}
which, after a shift $\ell\to \ell-k_{2}$ in the second integral, is exactly the same as Eq. (\ref{F380e}). These integrals vanish due to the traces.  We are therefore left with the potentially anomalous part of $\partial_{\lambda}\Gamma^{\mu\lambda\nu}\left(x,y,z\right)$, Eq. (\ref{F396ec}), which if one assumes gauge invariance can be set to zero. 
\par\vskip0.5cm
\noindent{\it ii) Axial Vector Ward Identity}
\par\noindent
Considering the three-point function (\ref{F373e}) and its derivative with respect to $x^{\mu}$, using the definition of time-ordered product we have again:
\begin{nedalph}\label{3100ea}
\partial_{\mu}\Gamma^{\mu\lambda\nu}\left(x,y,z\right) \equiv 
\big[\partial_{\mu}\Gamma^{\mu\lambda\nu}\left(x,y,z\right)\big]_{\mbox{\small{formal}}} +
\big[\partial_{\mu}\Gamma^{\mu\lambda\nu}\left(x,y,z\right)\big]_{\mbox{\small{anomal}}},
\end{eqnarray}
where  the formal part is given by:
\begin{eqnarray}\label{F3100eb}
\big[\partial_{\mu}\Gamma^{\mu\lambda\nu}\left(x,y,z\right)\big]_{\mbox{\small{formal}}} &\equiv& \delta\left(x^{0}-y^{0}\right)\bigg<T\left(\big[J^{0\left(5\right)}\left(x\right),J^{\lambda}\left(y\right)\big]J^{\nu}\left(z\right)\right)\bigg> \nonumber\\
&&\hspace{-0.5cm}+\delta\left(x^{0}-z^{0}\right)\bigg<T\left(J^{\lambda}\left(y\right)\big[J^{0\left(5\right)}\left(x\right),J^{\nu}\left(z\right)\big]\right)\bigg>,
\end{eqnarray}
and the anomalous part by: 
\begin{eqnarray}\label{F3100ec}
\big[\partial_{\mu}\Gamma^{\mu\lambda\nu}\left(x,y,z\right)\big]_{\mbox{\small{anomal}}} \equiv \bigg<\mbox{T}\left(\left(\partial_{\mu}J^{\mu\left(5\right)}\left(x\right)\right)J^{\lambda}\left(y\right)J^{\nu}\left(z\right)\right)\bigg>.
\end{nedalph}
The formal part can be evaluated using Eq. (\ref{F397ed}). Going through the same manipulations leading to the Eq. (\ref{F398e}) we obtain exactly the same integrals given in the Eq. (\ref{F390e}), which vanish. We therefore have:
\begin{eqnarray}\label{F3101e}
\big[\partial_{\mu}\Gamma^{\mu\lambda\nu}\left(x,y,z\right)\big]_{\mbox{\small{formal}}}=0,
\end{eqnarray}    
and are left with the anomalous part $\partial_{\mu}\Gamma^{\mu\lambda\nu}$, which we have already calculated, Eq. (\ref{F392e}), leading to the axial anomaly expression (\ref{F395e}).
\par\vskip0.5cm\par\noindent
\section{Conclusion}
\noindent
In this paper we have studied the Ward identities and calculated the axial anomaly in noncom\-mu\-ta\-tive (NC)-QED on {\bf R}$^{D}$. 
We started from the NC-QED action, originally given in Refs.\cite{hayakawa1,hayakawa2}. As in the case of the commutative non-Abelian gauge theory, the Ward identity in the simple case of fermion-antifermion annihilation into two gauge bosons is only satisfied, if the contribution of the three-gauge vertex to this process is also considered. In this calculation, we have made use of the usual dispersion relation, in contrast to Ref.\cite{susskind}, where matter in adjoint representation is used.
\par
We then derived the U(1)-anomaly in NC-QED using several methods carried over from the commutative field theory. First we studied the massless two-dimensional NC-QED. The axial anomaly was obtained, in the method of point-splitting after an integration over noncomutative space-time coordinates, which also preserved $\star$-gauge invariance of the axial anomaly.  Similar computation in four dimensions leads to the expected $\star$-gauge invariant result for the axial anomaly.    
\par
In the Fujikawa method, performing a gauge {\it covariant} regularization, the anomaly in  a gauge {\it covariant} form was obtained.  
Then a detailed analysis of the triangle diagrams for the axial vector current was carried out. The noncommutativity parameter $\theta_{\eta\sigma}$ appeared in the phases in the Feynman integrals for the triangle diagrams, but these phases were independent of the internal loop momentum. Hence no nonplanar diagrams appear at this level and known methods from commutative QED could be used to evaluate the vector and axial vector Ward identities in NC-QED. 
\par
We have shown that in momentum space the resulting expression for the axial anomaly differs from the corresponding commutative result by a factor $\cos\left(\frac{\theta_{\eta\sigma}}{2}k_{1}^{\eta}k_{2}^{\sigma}\right)$. Here $k_{1}$ and $k_{2}$ are the momenta of the gauge fields interacting with both vector currents. 
For non-vanishing parameter $\theta_{\eta\sigma}$, as in the commutative non-Abelian gauge theories, the anomalous contributions from the triangle diagrams do not yield the full anomaly and higher order diagrams have to be taken into account. These higher loop diagrams are also planar. This means that no UV/IR mixing will appear also in the higher loop order. This is certainly the consequence of the choice of the noncommutative U(1) current in the present paper. Further studies\cite{neda4} show that in noncommutative U(1) gauge theory with fundamental matter field coupling, there are in fact three different currents, which all, expressing the same global symmetry of the noncommutative action, lead to the same classically conserved charge. Two of them lead to planar triangle diagrams, whereas the third one includes planar as well as nonplanar diagrams, which have IR singularities. We study the effect of nonplanar diagrams to the gauge and global anomalies of  noncommutative U(1) and U(N) gauge theories in a second paper\cite{neda4}. For noncommutative U(1) gauge theory, it will turn out that for certain choice of noncommutativity parameter a finite global anomaly emerges from the nonplanar contributions to the triangle diagrams due to UV/IR mixing.       
\par
Finally, here,  a separation between the formal and the potentially anomalous part of the divergence of the vertex function was performed. This is analogous to the separation between the symmetric and antisymmetric part of traces over group generators in non-Abelian gauge theories\cite{weinberg}. Using the algebra of the NC-QED currents the formal parts of vector and axial vector were shown to vanish. Assuming gauge invariance of NC-QED the potentially anomalous part of the vector Ward identity vanish leaving the axial vector current anomalous.  
\par\vskip0.5cm\par\noindent
\nonumsection{Acknowledgments}\noindent
We would like to thank H. Arfaei for discussions and M. Hayakawa for a communication regarding Feynman diagrams.
\vskip0.5cm\noindent
{\bf Note added}: After this work (hep-th/0002143) other studies on the anomalies of non-Abelian noncommutative gauge theories appeared\cite{chiral}.  
\par\vskip1cm\noindent

\nonumsection{References}

\newpage
\nonumsection{Figures}
\begin{figure}[htbp]            
\SetScale{0.3}

\begin{picture}(90,60)(0,13)
\Text(20,5)[]{$p_{+}$}
\Text(90,5)[]{$p$}
\Text(20,60)[]{$k_{1},\mu$}
\Text(90,60)[]{$k_{2},\nu$}
\Text(60,-10)[]{$\left(1a\right)$}
\SetScale{0.8}
\LongArrow(35,48)(23,58)
\LongArrow(70,42)(60,42)
\LongArrow(102,20)(88,30)
\LongArrow(24,20)(37,30)
\LongArrow(90,48)(102,58)
\SetScale{0.3}
\Line(230,100)(100,100)
\Vertex(100,100){5}
\Vertex(230,100){5}
\Line(100,100)(0,20)
\Line(330,20)(230,100)
\Photon(230,100)(330,180){3}{6}
\Photon(100,100)(0,180){3}{6}
\end{picture}
\begin{picture}(70,60)(0,13)
\Text(40,30)[]{$+$}
\Text(80,5)[]{$p_{+}$}
\Text(150,5)[]{$p$}
\Text(70,60)[]{$k_{1},\mu$}
\Text(160,60)[]{$k_{2},\nu$}
\Text(115,-10)[]{$\left(1b\right)$}
\SetScale{0.8}
\LongArrow(130,70)(115,82)
\LongArrow(175,20)(162,30)
\LongArrow(140,42)(130,42)
\LongArrow(100,20)(113,30)
\LongArrow(148,70)(162,82)
\SetScale{0.3}
\Line(430,100)(300,100)
\Vertex(300,100){5}
\Vertex(430,100){5}
\Line(300,100)(200,20)
\Line(530,20)(430,100)
\Photon(300,100)(360,140){2}{4}
\Photon(380,150)(480,240){3}{6}
\Photon(430,100)(250,240){3}{8}
\end{picture}
\begin{picture}(70,60)(0,13)
\Text(110,30)[]{$+$}
\Text(135,5)[]{$p_{+}$}
\Text(220,5)[]{$p$}
\Text(150,60)[]{$k_{1},\mu$}
\Text(210,60)[]{$k_{2},\nu$}
\Text(180,-10)[]{$\left(1c\right)$}
\Text(200,40)[]{$k_{3},\rho$}
\Line(580,100)(480,20)
\Line(680,20)(580,100)
\Vertex(580,100){5}
\Photon(580,200)(580,100){3}{6}
\Photon(580,200)(480,240){3}{6}
\Photon(580,200)(680,240){3}{6}
\Vertex(580,200){5}
\SetScale{0.8}
\LongArrow(190,25)(203,35)
\LongArrow(243,25)(230,35)
\LongArrow(225,60)(225,50)
\LongArrow(227,85)(247,95)
\LongArrow(210,85)(190,95)
\end{picture}
\vskip1.5cm
\caption{Lowest order diagrams contributing to fermion-antifermion annihilation into a pair of gauge bosons. }
\end{figure}
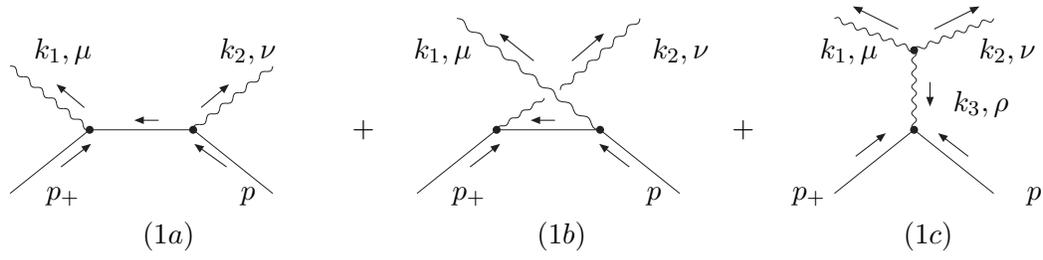

\parindent0em                  

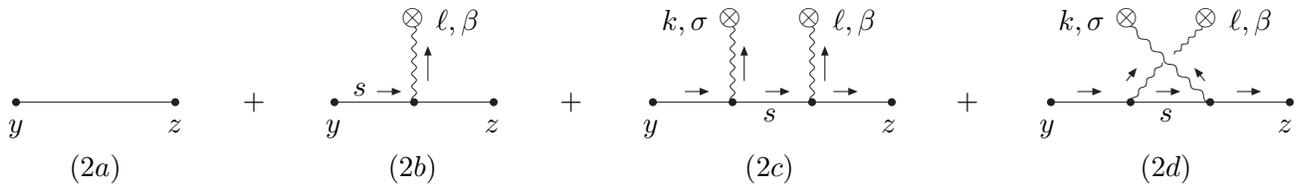
\begin{figure}[htbp]                              
\SetScale{0.3}
\begin{picture} (400,60)(0,20)
\Vertex(0,0){5}
\Vertex(200,0){5}
\Line(0,0)(200,0)
\Text(0,-10)[]{$y$}
\Text(60,-9)[]{$z$}
\Text(30,-25)[]{$\left(2a\right)$}
\Text(90,0)[]{$+$}
\Vertex(400,0){5}
\Vertex(600,0){5}
\Line(400,0)(500,0)
\Line(500,0)(600,0)
\Photon(500,0)(500,95){3}{6}
\Vertex(500,0){5}
\SetScale{0.8}
\LongArrow(194,10)(194,25)
\LongArrow(170,5)(180,5)
\SetScale{0.3}
\Text(120,-10)[]{$y$}
\Text(180,-9)[]{$z$}
\Text(150,32)[]{$\otimes$}
\Text(150,-25)[]{$\left(2b\right)$}
\Text(130,5)[]{$s$}
\Text(167,30)[]{$\ell,\beta$}
\Text(210,0)[]{$+$}
\Vertex(800,0){5}
\Vertex(1100,0){5}
\Text(240,-10)[]{$y$}
\Text(330,-9)[]{$z$}
\Text(285,-25)[]{$\left(2c\right)$}
\Text(253,30)[]{$k,\sigma$}
\Text(317,30)[]{$\ell,\beta$}
\Text(285,-5)[]{$s$}
\Line(800,0)(900,0)
\Line(900,0)(1000,0)
\Line(1000,0)(1100,0)
\Photon(900,0)(900,95){3}{6}
\Photon(1000,0)(1000,95){3}{6}
\Vertex(900,0){5}
\Vertex(1000,0){5}
\Text(270,32)[]{$\otimes$}
\Text(300,32)[]{$\otimes$}
\SetScale{0.8}
\LongArrow(381,10)(381,25)
\LongArrow(343,10)(343,25)
\LongArrow(315,5)(325,5)
\LongArrow(353,5)(363,5)
\LongArrow(385,5)(395,5)
\SetScale{0.3}
\Text(360,0)[]{$+$}
\Vertex(1300,0){5}
\Vertex(1600,0){5}
\Text(390,-10)[]{$y$}
\Text(480,-9)[]{$z$}
\Text(435,-25)[]{$\left(2d\right)$}
\Text(403,30)[]{$k,\sigma$}
\Text(467,30)[]{$\ell,\beta$}
\Text(435,-5)[]{$s$}
\Line(1300,0)(1400,0)
\Line(1400,0)(1500,0)
\Line(1500,0)(1600,0)
\Photon(1500,0)(1400,95){3}{6}
\Photon(1400,0)(1440,50){2}{4}
\Photon(1455,60)(1490,95){2}{4}
\Vertex(1400,0){5}
\Vertex(1500,0){5}
\Text(420,32)[]{$\otimes$}
\Text(450,32)[]{$\otimes$}
\SetScale{0.8}
\LongArrow(560,10)(556,15)
\LongArrow(523,10)(527,15)
\LongArrow(500,5)(510,5)
\LongArrow(537,5)(547,5)
\LongArrow(575,5)(585,5)
\end{picture}
\vskip1.5cm
\caption{Diagrams contributing to U(1)-anomaly in four dimensional massless NC-QED in the method of point-splitting.}
\end{figure}
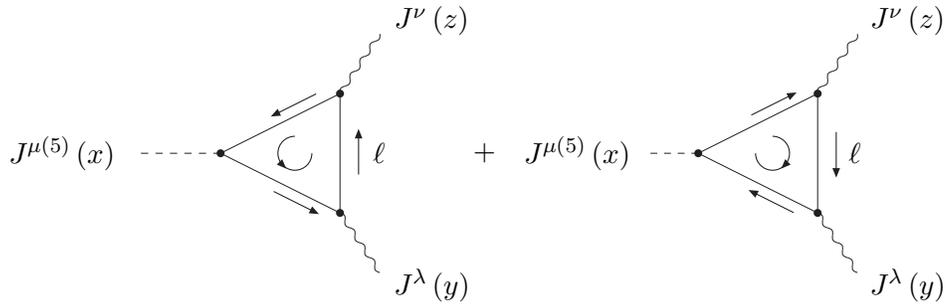
\begin{figure}[t]                              
\SetScale{0.3}
\begin{center}
\begin{picture} (200,40)(0,20)
\Vertex(0,0){5}
\Vertex(150,75){5}
\Vertex(150,-75){5}
\Line(0,0)(150,75)
\Line(0,0)(150,-75)
\Line(150,75)(150,-75)
\DashLine(-100,0)(0,0){10}
\Photon(150,75)(200,150){3}{4}
\Photon(150,-75)(200,-150){3}{4}
\Text(60,0)[]{$\ell$}
\Text(-60,0)[]{$J^{\mu\left(5\right)}\left(x\right)$}
\Text(80,50)[]{$J^{\nu}\left(z\right)$}
\Text(80,-50)[]{$J^{\lambda}\left(y\right)$}
\SetScale{0.8}
\LongArrow(65,-10)(65,10)
\LongArrow(45,28)(25,18)
\LongArrow(25,-18)(45,-28)
\ArrowArc(35,0)(8,90,360)
\SetScale{0.3}
\Text(100,0)[]{$+$}
\Vertex(600,0){5}
\Vertex(750,75){5}
\Vertex(750,-75){5}
\Line(600,0)(750,75)
\Line(600,0)(750,-75)
\Line(750,75)(750,-75)
\DashLine(540,0)(600,0){10}
\Photon(750,75)(800,150){3}{4}
\Photon(750,-75)(800,-150){3}{4}
\Text(240,0)[]{$\ell$}
\Text(135,0)[]{$J^{\mu\left(5\right)}\left(x\right)$}
\Text(260,50)[]{$J^{\nu}\left(z\right)$}
\Text(260,-50)[]{$J^{\lambda}\left(y\right)$}
\SetScale{0.8}
\LongArrow(290,10)(290,-10)
\LongArrow(250,18)(270,28)
\LongArrow (270,-28)(250,-18)
\ArrowArcn(260,0)(8,90,180)
\end{picture}
\end{center}
\vskip2.5cm
\caption{Triangle Diagrams for the anomaly in the axial vector current $J^{\mu\left(5\right)}\left(x\right)$ indicated by the dashed line.  }
\end{figure}
\begin{figure}[htbp]                              
\SetScale{0.3}
\begin{center}
\begin{picture} (350,60)(0,0)
\Vertex(0,0){5}
\Vertex(150,75){5}
\Vertex(150,-75){5}

\Line(0,0)(150,75)
\Line(0,0)(150,-75)
\Line(150,75)(150,-75)

\DashLine(-100,0)(0,0){10}
\Photon(150,75)(200,150){3}{4}
\Photon(150,-75)(200,-150){3}{4}
\Text(20,-60)[]{$\left(4a\right)$}
\Text(180,-60)[]{$\left(4b\right)$}
\Text(330,-60)[]{$\left(4c\right)$}
\Text(90,0)[]{$+$}
\Vertex(500,0){5}
\Vertex(650,0){5}
\Vertex(575,75){5}                                                                         
\Vertex(575,-75){5}
\DashLine(400,0)(500,0){10}
\Line(500,0)(575,75)
\Line(500,0)(575,-75)
\Line(650,0)(575,75)
\Line(650,0)(575,-75)
\Photon(575,75)(625,150){3}{5}
\Photon(575,-75)(625,-150){3}{5}
\Photon(650,0)(750,0){3}{5}
\Text(250,0)[]{$+$}
\Vertex(1000,0){5}
\Vertex(1075,75){5}
\Vertex(1075,-75){5}
\Vertex(1150,37.5){5}
\Vertex(1150,-37.5){5}
\DashLine(900,0)(1000,0){10}
\Line(1000,0)(1075,75)
\Line(1000,0)(1075,-75)
\Line(1150,37.5)(1075,75)
\Line(1150,-37.5)(1075,-75)
\Line(1150,37.5)(1150,-37.5)
\Photon(1075,75)(1125,150){3}{5}
\Photon(1075,-75)(1125,-150){3}{5}
\Photon(1150,37.5)(1225,80){3}{5}
\Photon(1150,-37.5)(1225,-80){3}{5}
\end{picture}
\end{center}
\vskip2.5cm
\caption{$\left(4a\right)$ Triangle diagram. $\left(4b\right)$ and $\left(4c\right)$ one-loop diagrams contributing to the anomaly in an axial vector current indicated by the dashed line.}
\end{figure}
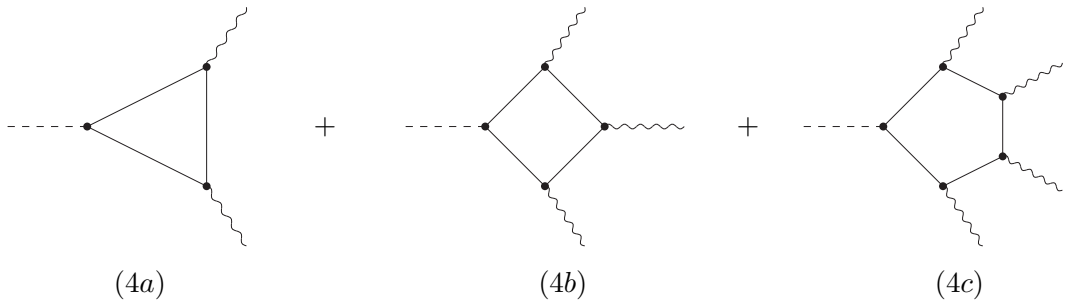

\end{document}